\documentclass[aps,pre,preprint,groupedaddress,floatfix]{revtex4-2}

\usepackage{graphicx}
\usepackage{amsmath,amssymb} 
\usepackage{color}
\usepackage{epstopdf}

\newcommand{\etal}{\textit{et al.}~}
\newcommand{\reffig}[1]{Fig.~\ref{#1}}
\newcommand{\eq}[1]{Eq.~\eqref{#1}}
\newcommand{\Rey}{\mathit{Re}}

\begin{document}
\title{Effect of viscosity on surface acoustic wave driven collective particle dynamics in sessile droplets: nebula, black holes and white dwarfs}
\author{Shuren Song}
\author{Jia Zhou} 
\email{jia.zhou@fudan.edu.cn}
\author{Antoine Riaud}
\email{antoine\_riaud@fudan.edu.cn}
\affiliation{\\State Key Laboratory of ASIC and System, \\School of Microelectronics, \\Fudan University, \\Shanghai, P. R. China}
\date{\today}
\begin{abstract}
Surface acoustic waves (SAW) can concentrate micro-particles in droplets within seconds. Yet, the mechanism is not clear and existing explanations fail by several orders of magnitude. In this paper, we analyze the effect of fluid viscosity and particle size on SAW-driven collective particle dynamics in droplets. In most of our experiments, the particles do not aggregate but instead remain away from the droplet center, thereby forming "black holes". We show that the black holes are due to steric hindrance wherein the poloidal streamlines that should drive particles to the center of the droplet come too close to the solid, so that the particles carried along these streamlines touch the solid wall on the edge of the black hole before reaching the center of the droplet. The size of these black holes is correlated with the size of the aggregates formed in less viscous droplets. This suggests a common formation mechanism for black holes and white dwarfs (aggregates). In the former, the particles touching the solid would be washed away by the fluid, whereas in the latter the particles would remain in contact with the solid and roll to the center of the droplet where an aggregate is formed. We also discuss the stability conditions of the aggregate at the bottom of the droplet. The concept of hydrodynamic shielding is then used to concentrate 1 $\mu$m particles using 10 $\mu$m beads as shields.
\end{abstract}
\maketitle

\section{Introduction}

\begin{figure*}
	\includegraphics[width=5in]{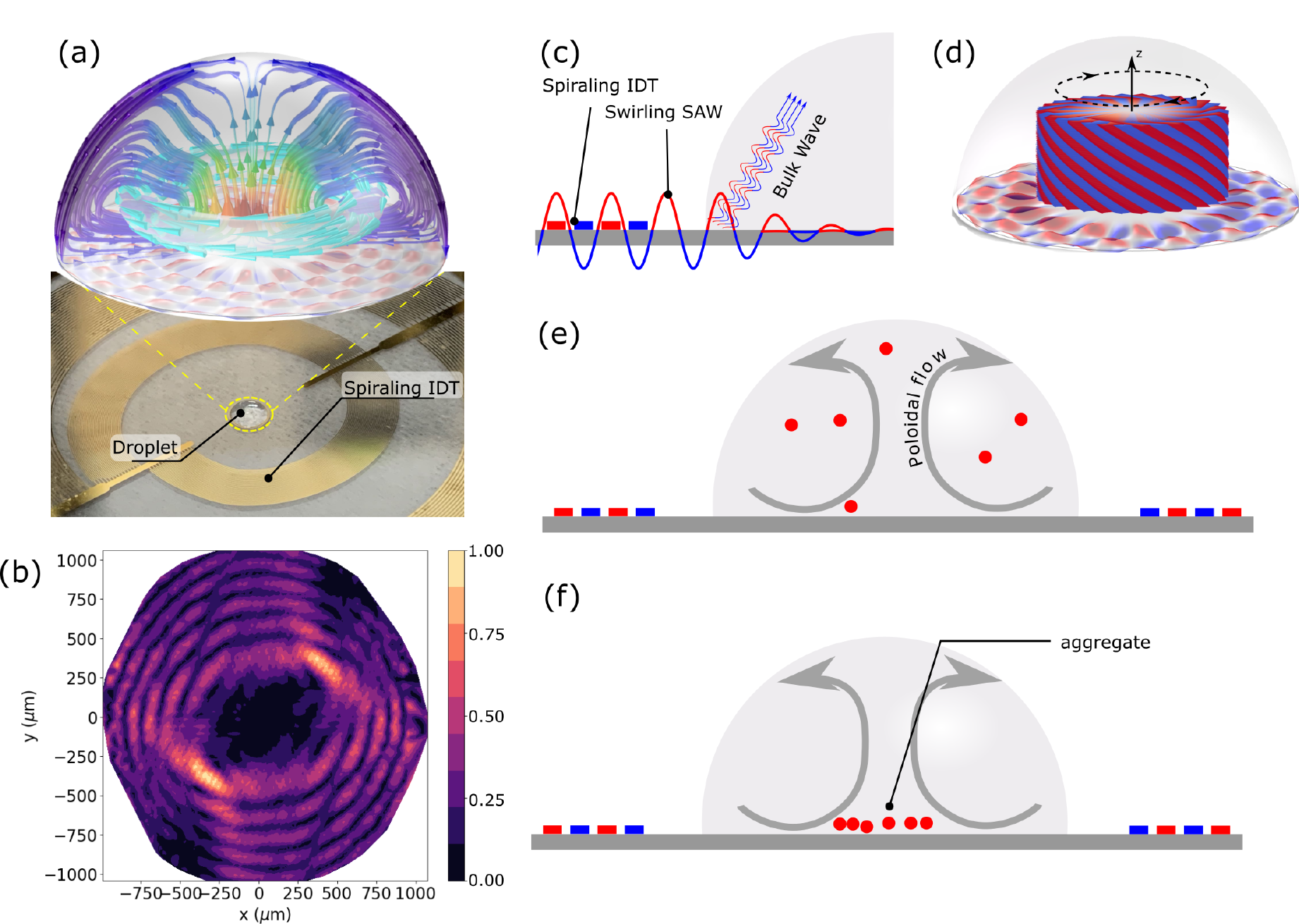}
	\caption{Experimental setup. (a) A droplet is placed at the center of a spiraling interdigitated transducer (IDT). When the IDT is turned on, the liquid in the droplet starts spinning and a poloidal flow is observed. (b) Experimental amplitude of the swirling surface acoustic wave (SAW) generated by the IDT. The dark region at the center indicates that the SAW has zero amplitude and guarantees the absence of electric forces acting on the particles in this region. The maximum SAW amplitude is nearly 1 nm. (c) Transmission of ultrasonic power from the IDT to the droplet. The spiraling IDT generates a swirling SAW ($\mathcal{W}_{-15}$) that radiates into the droplet as a bulk acoustic wave (acoustic vortex). (d)  Illustration of the incident swirling SAW (shown as an undulation at the bottom of the droplet) and the resulting three-dimensional acoustic vortex. The center of the vortex (in white) has zero-amplitude due to the topological protection, therefore ensuring the absence of acoustic radiation force acting on the particles. (e,f) Particle aggregation mechanism. (e) the bulk acoustic wave drives an acoustic streaming in the droplet with a poloidal component. (f) particles are carried along with the acoustic streaming and accumulate at a stagnation point at the base of the droplet. At this location, gravity is stronger than drag which pins the particles at the bottom.}
	\label{fig: setup}
\end{figure*}

\begin{figure*}[h!]
	\includegraphics[width=5in]{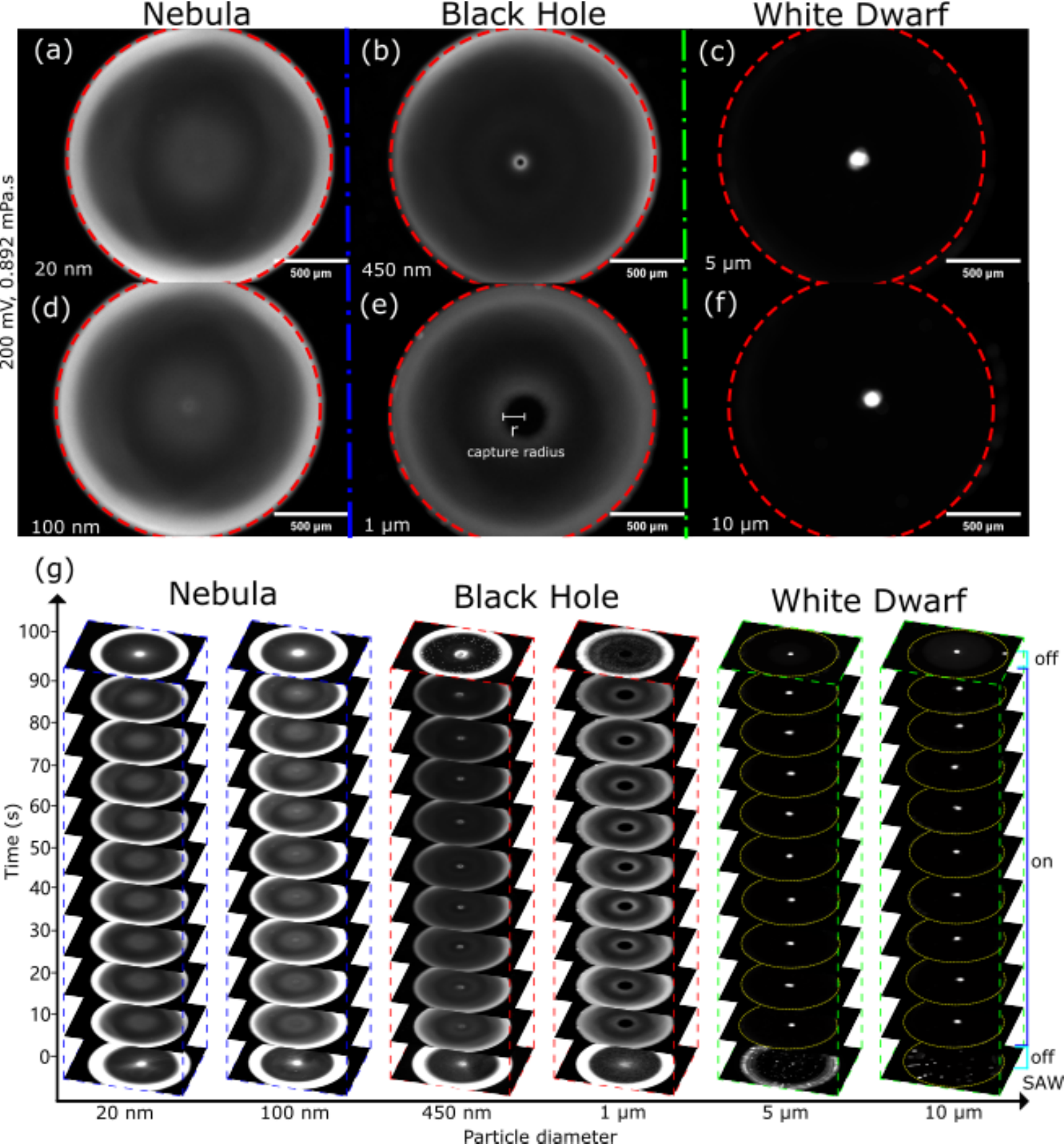}
	\caption{Three different outcomes of particle concentration in water droplets by an acoustic vortex depending on the particle diameter: (a,d) nebula (100 nm diameter particles), (b,e) black hole (1 $\mu$m diameter particles), (c,f) white dwarf (10 $\mu$m diameter particles). Each picture is the average of video frames obtained within the last 2 s before the SAW is turned off. (g) Evolution of the concentration patterns in the droplet. The photographs were taken as 10 s intervals. (Multimedia view)}
	\label{fig: particles_in_water}
\end{figure*}

Enrichment and separation of solid particles from microliter fluid samples is an essential preparatory step for labs-on-chip. Compared to other methods such as inertial microfluidics \cite{shelby2003high, chiu2007cellular, mach2011automated, martel2012inertial}, electrophoresis \cite{mampallil2013sample},  thermophoresis \cite{darhuber2005principles, hu2002evaporation, brutin2018recent} and acoustophoresis \cite{lenshof2011emerging,riaud2020mechanical,kolesnik2021unconventional}, the aggregation of particles in droplets using surface acoustic waves (SAW) is surprisingly fast (10 s) and simple to operate: a droplet is placed on the path of a SAW and starts spinning quickly when hit by the wave \reffig{fig: setup}(a). Particles are then observed to concentrate at the center of the droplet and form a dense aggregate, see \reffig{fig: particles_in_water} (c) and (f) (multimedia view).

Although the operation is straightforward, the mechanism for particle aggregation is not firmly established. From the earliest experiments of SAW-based droplet actuation, it was understood that the liquid flow in the droplet is due to acoustic streaming \cite{shiokawa1989liquid}, that is a steady hydrodynamic flow driven by the bulk acoustic wave radiating in the liquid as the SAW passes underneath the droplet \reffig{fig: setup}(c). The bulk wave gradually transfers its pseudo-momentum \cite{thomas2003pseudo} to the fluid while it is attenuated \cite{lighthill1978acoustic,eckart1948vortices}. The transferred pseudo-momentum then acts as an acoustic streaming force that drives the fluid motion. This mechanism has been confirmed by three-dimensional first-principle simulations \cite{riaud2017influence}. Yet, the relation between the flow motion and the aggregation is less clear.

In the earliest observations of SAW-driven particle concentration, Li \etal \cite{li2007surface} tentatively attributed the concentration to shear induced-migration. However, the authors themselves pointed out that this explanation requires unrealistically high concentration of particles. Raghavan \etal \cite{raghavan2010particle} then carefully reexamined the problem using particle image velocimetry and numerical simulations, and proposed that the particle aggregation was due to an analog of the tea-leaf effect (TLE) \cite{einstein1926cause}. In TLE, a poloidal flow circulation (observed experimentally by Raghavan \etal \cite{raghavan2010particle}) drives the particles from the edge of the droplet to the center \reffig{fig: setup}(e). The liquid then moves upward, however the particles are held back by their weight and remain trapped in the aggregate \reffig{fig: setup}(f). The effect of weight on trapping is also supported by experiments of Bourquin \etal \cite{bourquin2014rare} showing that only particles denser than the fluid form aggregates. Although the TLE hypothesis can qualitatively explain all the experimental observations of Raghavan and Bourquin \cite{raghavan2010particle,bourquin2014rare}, it is not very convincing for an order of magnitude analysis. For instance, the sedimentation speed of 10 $\mu$m particles is 3 $\mu$m/s, but the flow speed in the droplet is nearly 1000 times larger \cite{raghavan2010particle}. Accordingly, Peng \etal have simulated the aggregation of particles driven by a poloidal flow sufficiently slow, and found that aggregation was taking nearly 10 min instead of 40 s observed in their experiments. The discrepancy is blatant when considering that high-performance concentration devices proposed by Shilton \etal \cite{shilton2008particle} or Zhang \etal \cite{zhang2021microliter} can concentrate particles within 1 s. Hence, there is no consensus about tea leaf effect interpretation, and alternative aggregation mechanisms have been proposed from time to time. For instance Sudeepthi \etal \cite{sudeepthi2019aggregation} postulated that aggregation was driven by the hydrodynamic pressure force acting on the particles. The question of the aggregation mechanism also has implications on the range of particle size that can be concentrated. Experiments suggests that particles larger than the acoustic wavelength tend to be driven toward the edge of the droplet by the acoustic radiation force \cite{rogers2010exploitation,destgeer2016acoustofluidic,destgeer2017particle}, whereas sub-microscopic particles accumulate along the contact line \cite{destgeer2016acoustofluidic,gu2021acoustofluidic}. Throughout these previous studies, the effect of particle size \cite{li2007surface,rogers2010exploitation,destgeer2016acoustofluidic}, particle to fluid density ratio \cite{bourquin2014rare}, particle suspension solid fraction \cite{li2007surface}, particle to acoustic wavelength ratio \cite{destgeer2016acoustofluidic}, acoustic field \cite{shilton2008particle,zhang2021microliter}, ultrasonic power \cite{li2007surface,rogers2010exploitation} and droplet contact angle \cite{destgeer2017particle} have been analyzed. It is therefore frustrating that no quantitative law is available to predict if particle aggregation will occur, or what will be the size of the aggregate.

In this paper, we analyze the collective dynamics of particles in sessile droplets. Depending on the fluid composition and particle size, we observe three different dynamics visually reminiscent of astrophysics: the particles remain dispersed in the entire droplet (nebula), the particles are only observed outside a trapping radius (black holes) and the particles aggregate in a small dot (white dwarfs). Unlike previous studies, we take a holistic view of these phenomena and use the scaling laws of the black holes to shed light on the aggregation mechanism. Finally, we illustrate how this gained understanding enables aggregating smaller particles.

\section{Methods}

In order to simplify the analysis, we design the experiment to rule out the effects of the acoustic radiation force. This is done by using acoustic vortices instead of the plane waves used in most previous experiments \cite{zhang2009rapid, destgeer2016acoustofluidic}. These acoustic waves are also the first ultrasonic field to have enabled aggregation and recovery of the concentrated pellet \cite{zhang2021microliter}.  

Acoustic vortices (\reffig{fig: setup} (d)) are helical waves that rotate around their propagation axis. The pitch of the helix is called topological charge $\ell$ and is the ratio of the wave pseudo-orbital-angular-momentum and pseudo-momentum \cite{thomas2003pseudo}. Symmetry shows that the topological order is conserved when the wave propagates in an axisymmetric domain (such as the droplet). This places a condition on any acoustic field in the droplet akin to a diffraction limit wherein any acoustic field must rotate around the axis and can only exist outside a radius $r$ that satisfies $2\pi r \geq \ell\lambda$ (where $\lambda$ is the acoustic wavelength in water). This topological protection ensures that regardless of the scattering in the droplet, the acoustic field at the center of the droplet is null. Acoustic vortices can conveniently be synthesized from the radiation of swirling SAW \cite{riaud2015anisotropic,riaud2015taming} which can be synthesized using spiraling transducers \cite{riaud2017selective} \reffig{fig: setup}(a). In order to ensure that at least 50\% of the SAW power radiates in the droplet while not affecting too much the acoustic field in the solid, the actuation frequency is set to 20 MHz. The topological charge is set to $\ell$ = -15 ($\mathcal{W}_{-15}$ which ensures that no acoustic field exists within a radius of approximately 180 $\mu$m. Hence, any aggregation occurring within this region is guaranteed to be due to non-acoustic forces. Unless specified, a sinusoidal 200 mV$_{pp}$ (peak to peak) voltage, amplified by a radiofrequency amplifier (ZHL-5W-1+, minicircuits), is applied to the spiraling IDT to generate the acoustic streaming. The experimental surface wave amplitude is shown in \reffig{fig: setup}(b). The dark region at the center of the droplet \reffig{fig: setup}(b,d) indicates the absence of acoustic field in this region.

To vary the viscosity, we mix glycerol and water by various amounts. The viscosity and density are provided in Table \ref{tab: wat_glyc} (additional properties are available in ref. \cite{riaud2017influence}). We note that glycerol-water mixtures are denser than the polystyrene particles used here ($\rho_p = 1050$ kg/m$^3$), such that the particles cannot be trapped except in pure water according to Bourquin \etal experiments \cite{bourquin2014rare}. This enables the formation of black holes that have been helpful to understand the formation of particle aggregates.

\begin{table}
	\caption{Properties of water-glycerol mixtures (at 25$^o$C) \label{tab: wat_glyc}}
	\begin{ruledtabular}
		\begin{tabular}{ccc}
			mass fraction (w\%) & viscosity (mPa.s) & density (kg/m$^3$) \\
			0 & 0.89 & 1000 \\
			50 & 5.0 & 1130 \\
			70 & 18.1 & 1180 \\
			80 & 45.4 & 1210 \\
			90 & 156 & 1230
		\end{tabular}
	\end{ruledtabular}
\end{table}

\section{Results and discussions}

\subsection{Forces and orders of magnitude}
In most SAW-driven particle concentration, the particles are exposed to acoustic, electric, hydrodynamic and gravity forces. The drag and lift forces due to acoustic streaming are considered as hydrodynamic forces and will be discussed later. The remaining acoustic forces are mainly the acoustic radiation force \cite{baresch2013three,gong2021equivalence} and possibly some microstreaming \cite{doinikov1994acoustic,baasch2019acoustic}. They only exist if the particle is immersed in an acoustic field, and can be omitted at the center of the droplet thanks to the topological protection offered by the acoustic vortex. The electric forces (electrophoresis \cite{strobl2004carbon,ma2015patterning}) are due to the piezoelectric coupling with the substrate such that any surface acoustic wave field is coupled to an electric potential \cite{white1965direct,bleustein1968new}. However, the SAW field at the center of the droplet is null due to the phase singularity of the vortex, which ensures the absence of electric forces near the center of the droplet. Hence, only hydrodynamic forces remain.

In the following orders of magnitude estimates, we consider a polystyrene particle (diameter $d_p=10$ $\mu$m, density $\rho_p=1050$ kg/m$^3$) immersed in water (density $\rho_0=1000$ kg/m$^3$, viscosity $\mu=1$ mPa.s). The droplet has a diameter $R_d=1$ mm and is experimentally observed to rotate at a speed $\Omega\simeq100$ rad/s. The characteristic shear rate is taken as $\dot{\gamma} = \Omega$. When a relative velocity (slip velocity) between the particle and the fluid is needed to compute a force, it is assumed to be $\delta v\simeq \frac{R_d}{\tau} = 1$ mm/s with $\tau\simeq 1$ s the characteristic aggregation time. This is tantamount to assume that this force alone is responsible for particle aggregation. 

Because the particle is small, it moves nearly at the fluid velocity with a very small particle Reynolds number $\Rey_p = \frac{\rho_0 d_p\delta v}{2\mu}\simeq 2\times10^{-3}$. Accordingly, we first consider forces in the limit $\Rey_p=0$. Horizontal forces acting on the particles include the drag force (given in the bulk by Stokes law \cite{stone2000philip} $\mathbf{F}=3\pi d_p\mu\mathbf{\delta v}$, and near the walls by Goldman's law \cite{goldman1967slow} $\mathbf{F} = 5.1\pi\mu{d_p}^2\dot{\gamma}$).  The particle is also exposed to a combination of accelerations and pressure forces that counteract each-other. In the vertical direction, the acceleration of gravity is balanced by buoyancy ($\mathbf{F} = (m_p-m_f)\mathbf{g}$) whereas in the radial direction the centrifugal acceleration is balanced by the radial pressure distribution ($\mathbf{F} = (m_p-m_f)\mathbf{g'}$). The fastest rotating water droplets are most likely to exhibit the strongest centrifugal forces. Their droplet Reynolds number $\Rey = \frac{\rho_0 {R_d}^2 \Omega}{\mu}$ approaches 100, which allows to express the acceleration $\mathbf{g'}$ in the inviscid approximation  $\mathbf{g'} = \mathbf{v}\cdot\nabla\mathbf{v}\simeq \frac{{v_\theta}^2}{r}\mathbf{e}_r$. We will subsequently see that centrifugal forces are negligible even in this case.

At nonvanishing particle Reynolds number, the particle also experiences lift, which is important in shear-induced migration. Saffman lift \cite{stone2000philip} ($\Rey_p \ll {\Rey_\gamma}^{1/2} \simeq 0.05 \ll 1$ and particle slip parallel to the flow direction) scales as $F = 3.23d_p\mu\Rey_\gamma^{1/2}\delta v$, where $\Rey_{\gamma} = \frac{\rho_0\dot{\gamma}{d_p}^2}{4\mu}$ is the shear Reynolds number of the particle. The lift of a particle in contact with a wall was studied by Krishnan \etal \cite{krishnan1995inertial}. Expressions and resulting numerical values of forces acting on the particles in the bulk and at the bottom of the droplet are summarized in Tables \ref{tab: mig_speed} and \ref{tab: vertic_F}.

\begin{table}
	\caption{Radial migration speed \label{tab: mig_speed}}
	\begin{ruledtabular}
		\begin{tabular}{lcc}
			Force & Expression & Migration velocity \\
			Centrifuge and pressure & $(m_p-m_f) R_d \Omega^2$ & 3  $\mu$m/s\\
			Lift \cite{stone2000philip} & $3.23d_p\mu\Rey_\gamma^{1/2}\delta v$ & 20 $\mu$m/s\\
			Drag (bulk) & $3\pi d_p \mu \delta v$  & 1 mm/s\\
			Drag (surface) \cite{goldman1967slow} & $3\pi {d_p}^2 \mu \dot{\gamma}$  &  1 mm/s\\
		\end{tabular}
	\end{ruledtabular}
\end{table}
According to Table \ref{tab: mig_speed}, the motion due to drag is much faster than the one due to any other force. In addition, there is a mismatch between the lift velocity (20 $\mu$m/s) and the drift velocity ($\delta v\simeq 1$ mm/s) that was originally assumed to calculate this lift velocity. Therefore, lift is not responsible for the migration of particles to the center of the droplet. This supports the TLE model where the in-plane motion of the tea leaves (particles) is solely due to drag.

\begin{table}
	\caption{Vertical forces \label{tab: vertic_F}}
	\begin{ruledtabular}
		\begin{tabular}{lcc}
			Force & Expression & Magnitude \\
			Weight and buoyancy & $(m_p - m_f) g$ & 0.3 pN\\
			Drag (bulk) & $3\pi d_p \mu \delta v$  & 100 pN\\
			Lift (stuck sphere) \cite{krishnan1995inertial} & $9.2\frac{\mu^2}{\rho_0}{\Rey_\gamma}^2$ & 0.06 pN\\ 
			Lift (free rolling sphere) \cite{krishnan1995inertial} & $1.7\frac{\mu^2}{\rho_0}{\Rey_\gamma}^2$ & 0.01 pN 
		\end{tabular}
	\end{ruledtabular}
\end{table}

Table \ref{tab: vertic_F} shows the forces in the vertical direction. It indicates that drag is considerably larger than gravity forces (weight and buoyancy), such that the vertical motion of the particles in the bulk is likely to go along with the flow. However, drag is parallel to the walls at the bottom of the droplet (and therefore has no vertical component). Thus, only lift exists when particles are in contact with the solid, and gravitational forces are larger than lift, which suggests that contact with the floor is stable.

\subsection{Three different concentration outcomes} 

\begin{figure*}[h!]
	\includegraphics[width=5in]{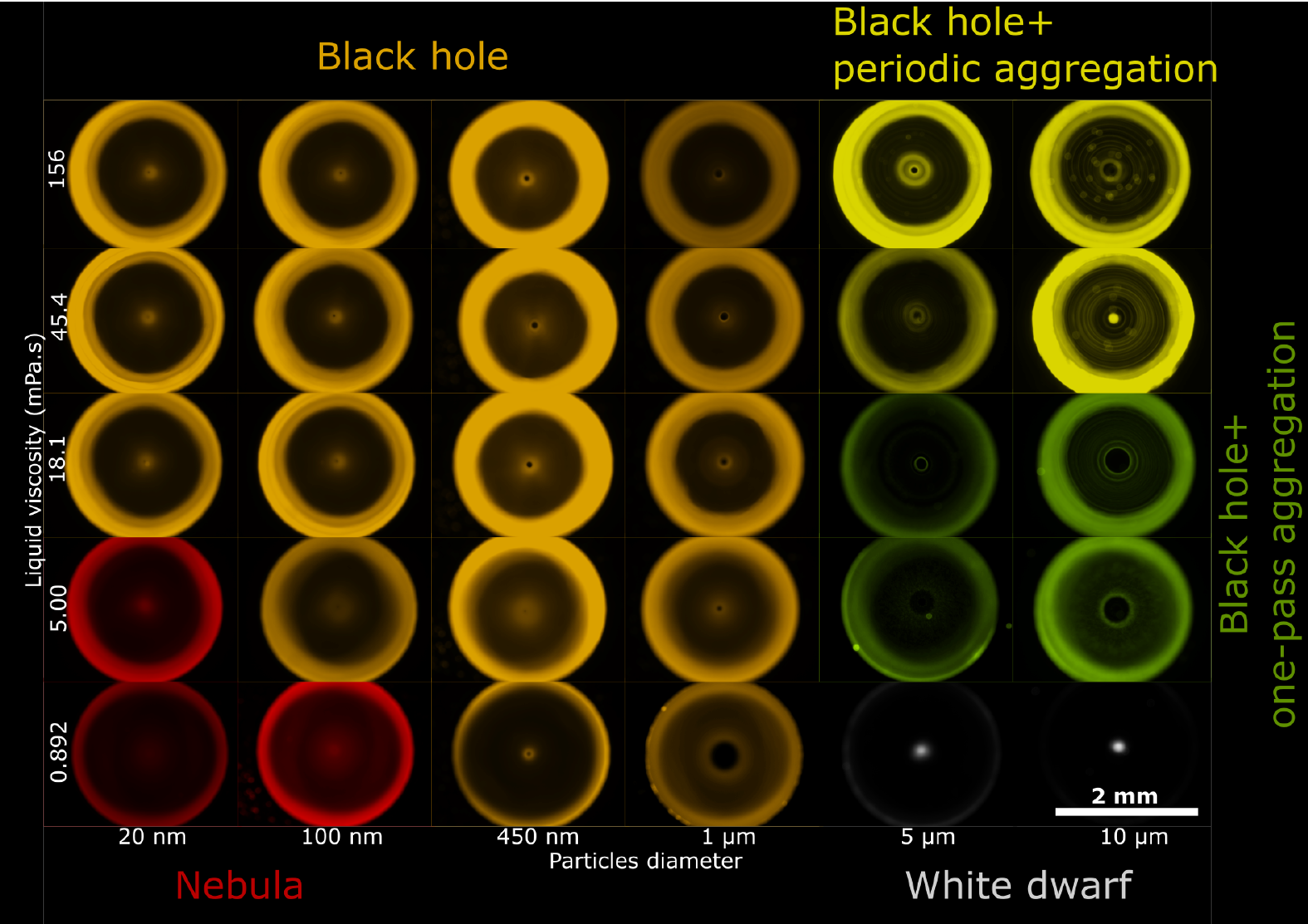}
	\caption{Three different outcomes of particle concentration in water-glycerol mixture droplets by an acoustic vortex depending on the particle diameter. Each picture is the average of video frames obtained within the last 4 s before the SAW is turned off. The outcome is color-coded, with nebula shown in red (lower left), black holes in orange (upper left), white dwarfs in white (lower right). More complex combinations include black holes and unstable aggregate that appear only once (in green, middle-right) or appear periodically (in yellow, upper right).}
	\label{fig: particles_in_different liquid}
\end{figure*}

The dynamics of particles with different diameters ($d_p$ = 20 nm, 100 nm, 450 nm, 1 $\mu$m, 5 $\mu$m, 10 $\mu$m) dispersed in water are shown in \reffig{fig: particles_in_water}. Three types of pattern are observed: nebula  (a,d) black hole (b,e) and white dwarf (aggregate, e,f). These patterns are generated within 10 s of ultrasound exposure and remain stable during the entire 100 s of the experiment, as attested by experimental pictures taken every 10 s (g).

In order to clarify the origin of these patterns, we alter the force balance by using particles of various sizes (20 nm, 100 nm, 450 nm, 1 $\mu$m, 5 $\mu$m, 10 $\mu$m) and using water-glycerol mixtures of increasing density and viscosity ($\mu$ = 0.892 mPa.s, 5 mPa.s, 18.1 mPa.s, 45.4 mPa.s, 156 mPa.s). \reffig{fig: particles_in_different liquid} displays the average of the last 4 s of the videos.

Over a range of viscosities and particle sizes, the black holes are the most common patterns found in our experiments. Aggregation is observed in all experiments using 5 and 10 $\mu$m diameter particles, but the aggregates are only stable in water. Black holes position themselves at an intermediate level between pattern-free nebula and the aggregates, and therefore may provide a glimpse on the dynamics of aggregation of particles in droplets. 

\subsection{Formation mechanism of the hydrodynamic black holes} 

\begin{figure*}[h!]
	\includegraphics[width=5in]{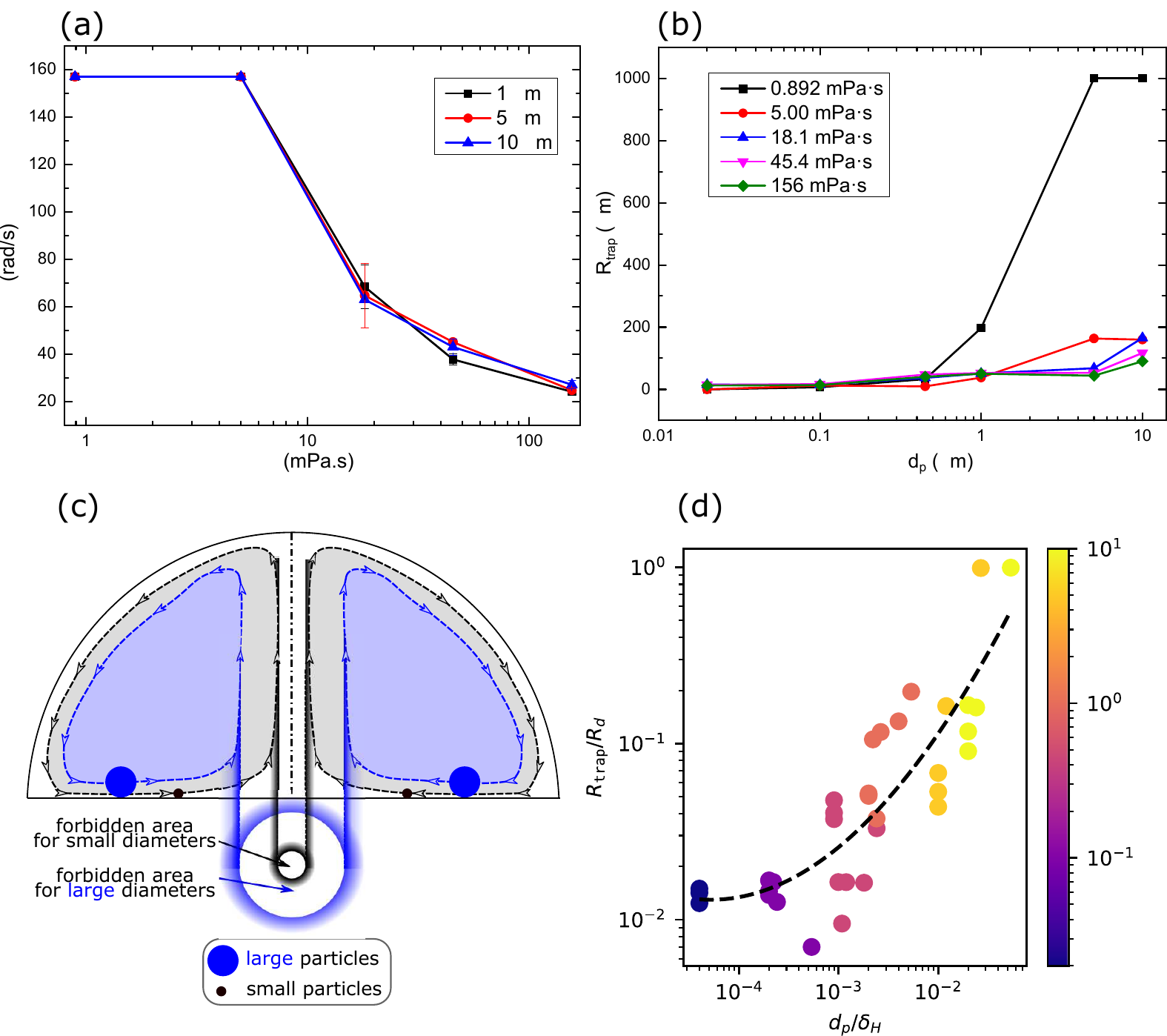}
	\caption{(a) Droplet rotation speed $\Omega$ depending on the liquid viscosity and particle size. (b) Trapping radius of the black holes depending on the particle diameter for different liquid viscosities. (c) Proposed generation mechanism for the black holes. Unlike small particles, large particles are unable to flow along the streamlines that go closest to the solid substrate, and therefore cannot reach the center of the droplet, which results in a forbidden area. At low viscosity (high droplet Reynolds number), the streamlines at the bottom of the droplet are squeezed into a thin boundary layer, such that large particles are prevented to access a much larger area. (d) Trapping radius of the black hole depending on the ratio between the particle diameter and the hydrodynamic layer thickness (the color indicates the particle diameter in $\mu$m).}
	\label{fig: dark holes}
\end{figure*}

When a black hole is formed in the droplet, no particle is observed within a trapping radius shown in \ref{fig: dark holes}(b). This trapping radius increases with the particle size and decreases with the viscosity.  When the particle diameter is small (20 nm and 100 nm), the trapping radius (that is, the radius of the dark hole) approaches zero, which can be regarded as a nebula. Because the black holes continue to exist even at very high viscosities, our force balance analysis suggests that they are formed via a purely steric mechanism. We postulate that particles are carried to the center of the droplet via some poloidal streamlines (see \reffig{fig: dark holes} (c)). The closer the streamline to the axis of the droplet, the closer it also flows next to the solid substrate. Therefore, large particles cannot flow along the streamlines closest to the center of the droplet, which results in large black holes for 5 and 10 $\mu$m diameter particles in high viscosity droplets ($\mu> 5$ mPa.s). In lower viscosity droplets, such as pure water, the holes tend to become larger. This may be because the flow recirculation is now confined to a thin boundary layer at the bottom of the droplet. Indeed, the droplet Reynolds number approaches 100, and therefore the flow is essentially inviscid with a thin viscous boundary layer near the solid surface. The particle size should then be compared to the thickness of this boundary layer. In order to consider both high and low viscosities, we estimate the boundary layer thickness as the minimum between half the droplet radius and the Blasius boundary layer thickness $\delta_H$, with the characteristic distance taken as half the droplet radius:

\begin{equation}
	\delta_H \approx \min\left(\frac{R_d}{2},5.0\sqrt{\frac{\nu R_d}{2 v}}\right), \label{eq3: Delta}
\end{equation}

The flow velocity  $v$ in \eq{eq3: Delta} is estimated as $v = Rd\Omega$ at different viscosities based on experimental measurements of the droplet rotation speed $\Omega$ (measured at the center of the droplet) (\reffig{fig: dark holes} (a)). 

If the black holes are truly formed due to the contact between the particles and the solid surface, then the trapping radius of the black holes should only depend on the droplet radius and the ratio of the particle size to the boundary layer thickness. \reffig{fig: dark holes} (d) compares the trapping radius of the black holes to the ratio between the particle size and the boundary layer thickness as given by \eq{eq3: Delta}. The clean relationship supports the idea that black holes are due to the collision of the particles with the substrate before they reach the center of the droplet.

\subsection{Implications for particles aggregation}

\begin{figure*}[h!]
	\includegraphics[width=5in]{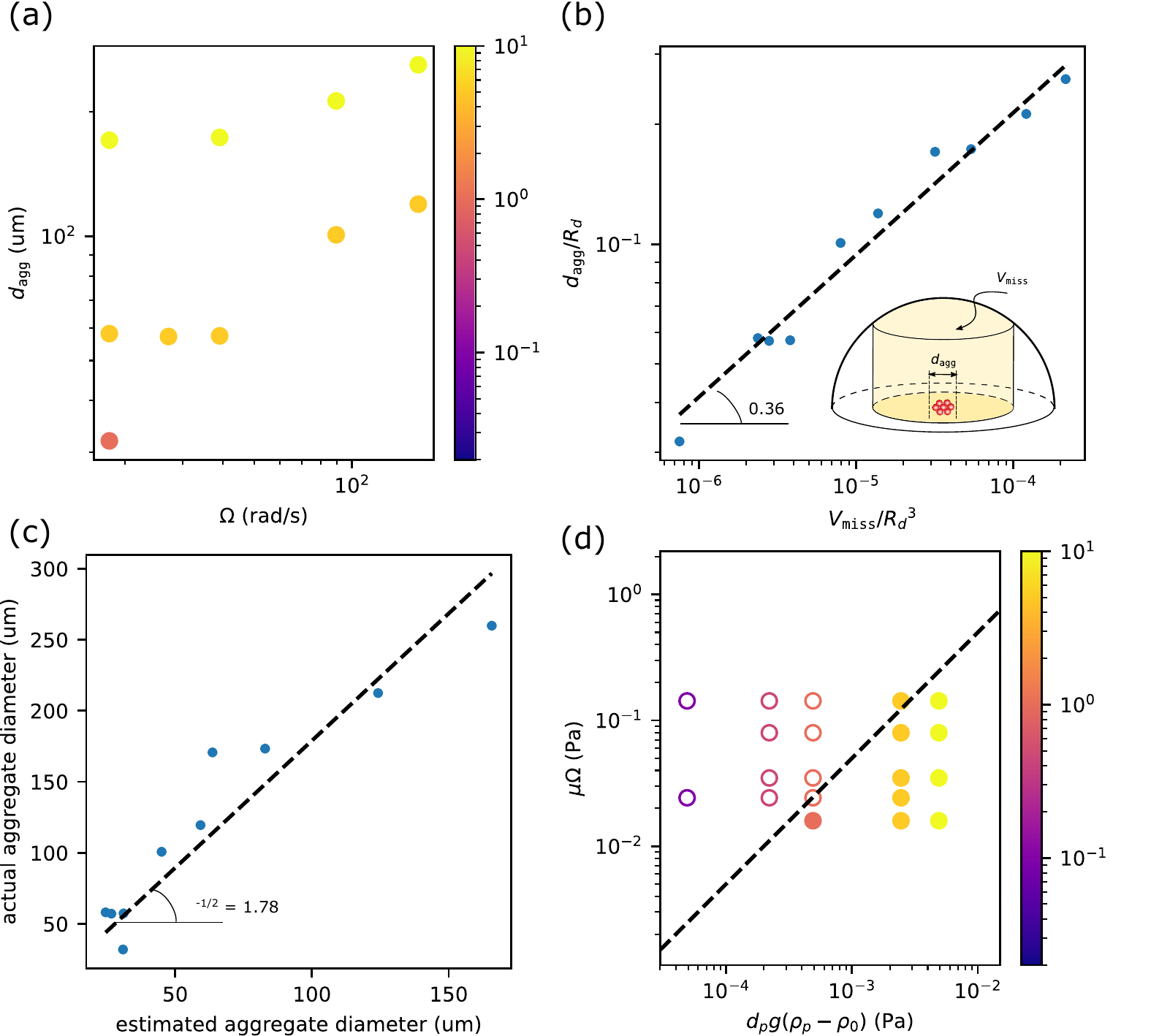}
	\caption{(a) Aggregate size depending on the rotation speed of the droplet (the color indicates the particle diameter in $\mu$m). (b) Aggregate diameter depending on the missing volume. (c) Comparison of the actual and estimated diameter assuming the the mising particles aggregate to form a disc-shaped monolayer. (d) Aggregate stability depending on the destabilizing shear stress and stabilizing hydrostatic pressure (the color indicates the particle diameter in $\mu$m).}
	\label{fig: aggregate_size}
\end{figure*}

The existence of black holes suggest that some amount of particles have been removed from the flow circulation in the droplet. The size of the aggregates follows a similar trend with the black holes trapping radius: it increases with the particle size and with the droplet rotation speed (\reffig{fig: aggregate_size}(a)). Therefore, we use the previously developed model to extrapolate the back hole sizes to the case where aggregates are formed, deduce the volumes $V_\text{miss}$ of the missing particles and compare it to the size of the aggregates obtained with 1, 5 and 10 $\mu$m particles in water using different power levels (\reffig{fig: aggregate_size}(b)). We find a clear correlation $d_\text{agg,exp}/R_d = 5.72 {V_\text{miss}}^{0.36}/R_d^3$ between the experimental aggregate diameter $d_\text{agg,exp}$ and the missing volume which suggests that the missing particles form the aggregate. 

Although the nearly $1/3$ exponent suggests that the particles form a three-dimensional aggregate, we find that a monolayer disc of volume $V_\mathtt{miss}$, height $d_p$ and diameter $d_\text{agg,est}$ describes better the experimental aggregate diameter $d_\mathtt{agg,exp}$ than a sphere ((\reffig{fig: aggregate_size}(c)). Indeed, the disc compacity is found to be $\phi = \frac{\pi {d_\text{agg,exp}}^2 d_p/4}{\pi {d_\text{agg,est}}^2 d_p/4} = 0.31$, close to the maximum compacity of a monolayer of spheres ($\frac{\pi}{3\sqrt{3}}\simeq 0.60$) and also more reasonable than a compacity of 0.02 if assuming a spherical aggregate. Therefore, the formation of the aggregates would proceed through the following steps: (i) the acoustic streaming drives a poloidal flow motion, (ii) the particles driven by the flow hit the substrate due to their finite size, and start rolling, (iii) they roll until reaching the center of the droplet, where they accumulate. The stability of the aggregate is considered in the next section.

\subsection{Stability of the aggregate}

 \begin{figure*}[h!]
	\includegraphics[width=5in]{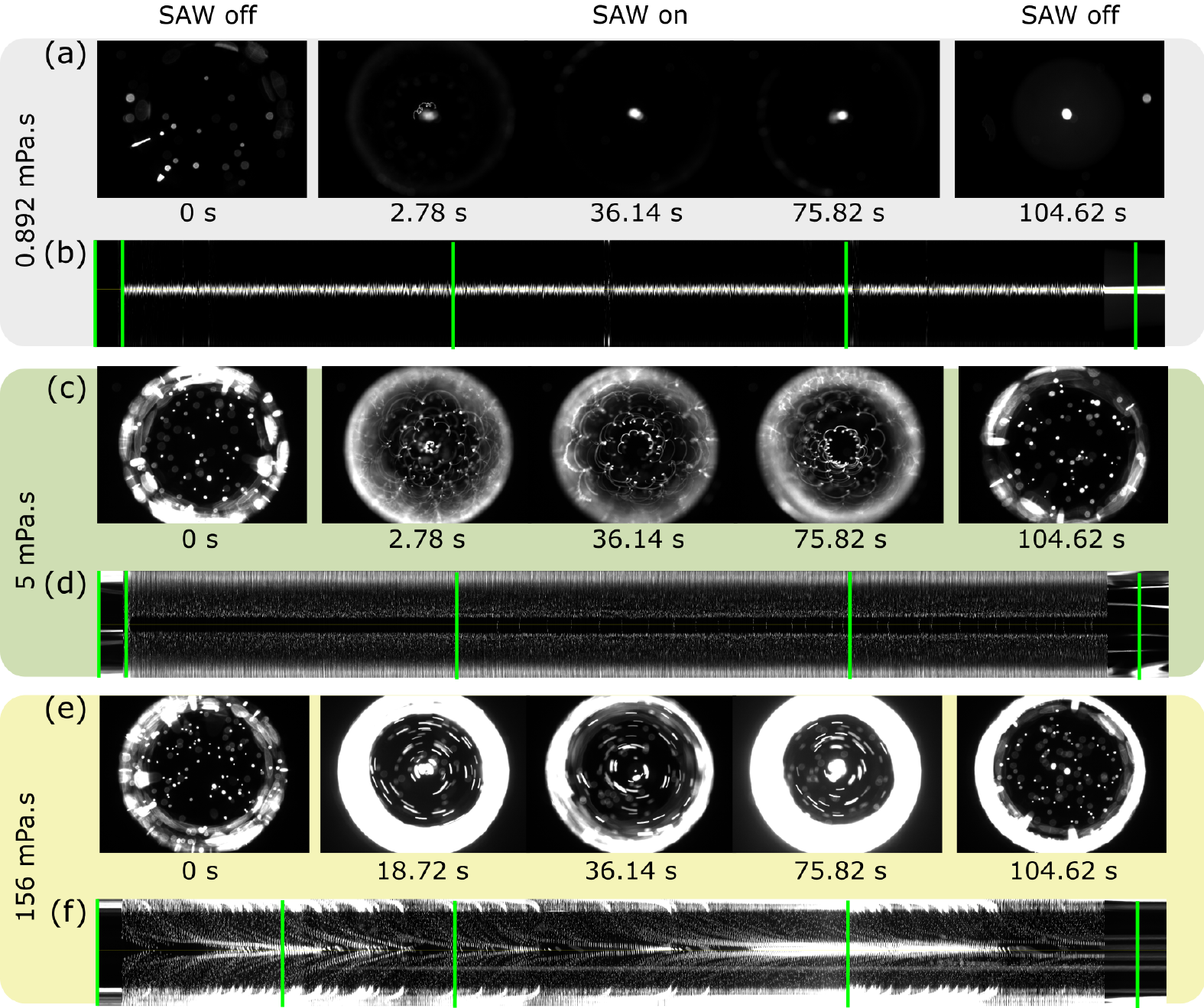}	
	\caption{Particles concentration dynamics of 10 $\mu$m particles in droplets of increasing viscosities. (a,c,e): successive pictures; (b,d,f): spatio-temporal projections. The green lines indicate the times of the photographs shown above. (a,b) stable aggregates formed in water droplets; (c,d) unstable aggregate breaking up immediately after forming in 50 w\% water-glycerol mixture; (e,f) unstable aggregate forming and breaking up multiple times in concentrated glycerol solution (90 w\%).}
	\label{fig: different_viscosity_5 um}
\end{figure*}

In the proposed black-hole model, the size of the aggregate is set by the ratio of the particle size to the viscous boundary layer at the bottom of the droplet. However, experiments in water-glycerol mixtures feature unstable aggregates that eventually break up. We unravel the phenomenon in a spatio-temporal cross-section of the droplet (\reffig{fig: different_viscosity_5 um}). Three distinct dynamics are observed with increasing glycerol fraction: (i) the aggregate is stable (pure water), (ii) the aggregate forms once and disappears permanently (iii)  the aggregate is formed and broken up regularly. Assuming that the cycles of aggregates creation/destruction are due to the the poloidal flow that washes particles and periodically brings them back, we conclude that the aggregate is formed during the first pass of the poloidal flow, as opposed to a gradual accumulation that would have been expected under the action of radial forces. This one-pass process explains the very fast aggregation speed observed experimentally \cite{shilton2008particle}. The distinction between aggregates that form only once and those to reappear periodically may be due to different dynamics of particles during the recirculation near the droplet contact line, which is beyond the scope of this paper. Neveretheless, we note that all our water-glycerol mixtures are denser than the particles, which forbids trapping according to Bourquin \etal \cite{bourquin2014rare}. Although this explains why particles are never trapped in water-glycerol mixtures, it gives no indication about the power levels or particle size that would allow trapping. Indeed, our experiments using water droplets exposed to various ultrasonic power show there is a minimum particle size that can be aggregated, and that this size depends on the rotation speed of the droplet (\reffig{fig: aggregate_size}(d)), in agreement with Li \etal \cite{li2007surface}.

To estimate the stability of the aggregate, we look at the stability of sphere aggregates in contact with a wall exposed to shear. Two cases may be distinguished (i) an isolated sphere exposed to a Couette flow \cite{goldman1967slow,krishnan1995inertial}, (ii) a sphere positioned atop of a regular monolayer of other spheres, and exposed to a Couette flow \cite{shields1936anwendung,agudo2017shear}. For isolated spheres, lift \cite{krishnan1995inertial} is insufficient to balance the weight of the spheres, and therefore spheres in contact with the walls should remain stable. However, under the effect of drag, a sphere monolayer may be expected to buckle, and the emerging spheres may be dragged away by the flow \cite{agudo2017shear}. Agudo \etal \cite{agudo2017shear} have shown that particles located on a monolayer of sphere could be dislodged when the torque applied by the drag force exceeds that of gravity, which depends on the dimensionless Shields number \cite{shields1936anwendung} $\theta = \frac{\mu\dot{\gamma}}{(\rho_p-\rho_0)gd_p} = \frac{\mu\Omega}{(\rho_p-\rho_0)gd_p}$. This number represents the ratio of drag to buoyancy. We note that, due to the no-slip boundary conditions on the wall, the slip velocity in the drag force has to be replaced by the shear, which yields $F=2.55\pi\mu d_p^2\dot{\gamma}$ \cite{goldman1967slow}. This scaling differs from previous studies \cite{raghavan2010particle,bourquin2014rare} by a factor $0.85 d_p/R_d\simeq 10^{-2}$. It therefore becomes more reasonable to compare this drag force to the particles weight. The Shields number can be interpreted as the ratio of viscous to gravitational torque $\frac{\mu\dot{\gamma}d_p^3}{(\rho_p-\rho_0)gd_p^4}$, the ratio of drag to weight $\frac{\mu\dot{\gamma}d_p^2}{(\rho_p-\rho_0)gd_p^3}$ or the ratio of shear stress to hydrostatic pressure. The latter interpretation is advantageous as it the least skewed by $d_p$ and therefore allows distinguishing other effects such as the shear rate. The state of each aggregate observed in our experiments is presented in \reffig{fig: aggregate_size}(d). We find that those states can be separated by a line $\mu\dot{\gamma} =\mu\Omega = \theta_c(\rho_p-\rho_0)gd_p$, with $\theta_c = 0.05$, in agreement with theoretical and experimental studies that report $0.01<\theta_c<0.1$ depending on the flow conditions and the pattern of the spheres \cite{agudo2017shear,topic2019effect}. 

\subsection{particle concentration shock wave}
While the shield is necessary to maintain the stability of the aggregate, the creation of the shield itself is an important question. We recall that the aggregate is formed by rolling particles. The speed $v_r(R)$ of a rolling sphere is given by Goldman \cite{goldman1967slow} as $v_r = \frac{1}{2}d_p\dot{\gamma}\eta$ with $\eta = \frac{0.74}{0.50-0.2\ln(\delta/d_p)}$ the ratio of the speed between a particle that would be freely floating (without wall effects) and a rolling particle. $\delta$ is a cutoff size where molecular forces dominate over hydrodynamic effects (between 1 nm and 0.1 nm). Accordingly, $0.26 < \eta < 0.40$ for particle diameters ranging from 1 $\mu$m to 10 $\mu$m. This indicates that rolling particles move more slowly than floating ones (\reffig{fig: separation}(a)). A concentration front is therefore formed between the slow-moving particle near the center of the droplet and fast-incoming floating particles on the periphery of the droplet. It is likely that dense layers of rolling particles move even more slowly due to hydrodynamic interactions between them. This creates a shock of highly concentrated particles converging to the center of the droplet, which has been observed in our experiments (\ref{fig: separation}(b)).

\subsection{Seed-assisted aggregation}

\begin{figure*}
	\centering
	\includegraphics[width=5in]{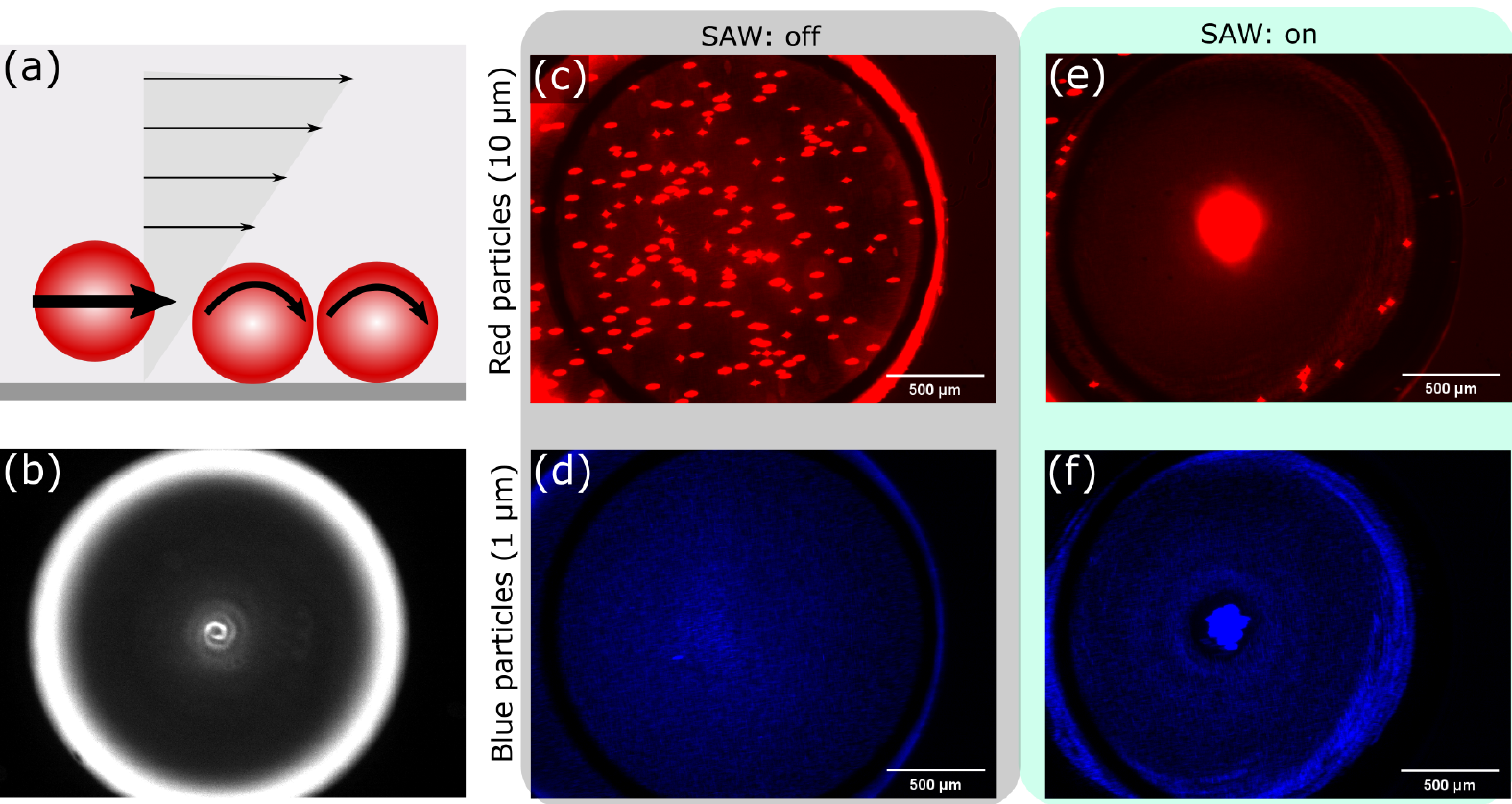}
	\caption{Collective particle dynamics in SAW-driven droplets. (a) particle concentration shock wave: rolling particles move slower than incoming floating particles, which creates a high-concentration front. Spiraling concentration shock wave of 1 $\mu$ particles (b). (c-f) Shields effect \cite{shields1936anwendung}: large particles (10 $\mu$, red fluorescence) aggregate at the center of droplet and protect smaller particles (1 $\mu$, blue fluorescence) from the drag of the acoustic streaming. Therefore, small particles can aggregate at the center of the droplet, which is not possible without large particles.}
	\label{fig: separation}
\end{figure*}

According to the proposed model, the stability of the aggregate depends strongly on the shielding of aggregated particles from hydrodynamic forces. It is therefore possible to shift the balance between drag and gravity by creating an aggregate with large particles and then use it the shielding effect to aggregate smaller ones. This is shown in \reffig{fig: separation} where 1 $\mu$m diameter particles (blue) can be aggregated inside a cluster of  10 $\mu$m dimeter particles (red) that shields them from the flow, even though 1 $\mu$m particles alone could not be aggregated. 

\section{Conclusion}
In order to understand better the SAW-driven aggregation of particles in sessile droplets, we have varied the viscosity of the liquid. We have found that particles tend to organize as hydrodynamic black holes where the center of the droplet is empty of particles. The geometry of these black holes is dictated purely by geometric constraints where the largest particles are physically forbidden to reach the center of the droplet as they touch the solid substrate before doing so. When the particles are lighter than the liquid, they gradually detach from the solid and escape by buoyancy, whereas particles heavier than the solid continue to roll toward the center of the droplet. When the particles concentration is high enough, slowly rolling particles are caught up by the floating ones, which creates a shock wave of highly concentrated particles. Finally, the aggregate stability is found to depend on the balance between buoyancy and drag as mentioned by previous studies, and discrepancies in terms of order of magnitude are addressed once accounting for the shielding effect of the solid wall and nearby particles. This shielding effect allows the concentration of smaller particles by using larger ones as shields, which has not been demonstrated previously. Finally, this paper has largely focused on particle dynamics in the droplet bulk, but some exciting development has shown that the smallest particles accumulate instead along the contact line of the droplet \cite{destgeer2016acoustofluidic} and can be separated using droplet doublets \cite{gu2021acoustofluidic}. Recent theoretical progress showing the enhancement of acoustic radiation pressure near sharp edges \cite{doinikov2020acoustic} suggests that this is a different concentration mechanism that should also be investigated in order to gain a full understanding on particle concentration in SAW-driven droplets. 

\section{Data Availability}
The data that support the findings of this study are available from the corresponding authors upon reasonable request.

\bibliography{References}

\begin{thebibliography}{46}%
\makeatletter
\providecommand \@ifxundefined [1]{%
 \@ifx{#1\undefined}
}%
\providecommand \@ifnum [1]{%
 \ifnum #1\expandafter \@firstoftwo
 \else \expandafter \@secondoftwo
 \fi
}%
\providecommand \@ifx [1]{%
 \ifx #1\expandafter \@firstoftwo
 \else \expandafter \@secondoftwo
 \fi
}%
\providecommand \natexlab [1]{#1}%
\providecommand \enquote  [1]{``#1''}%
\providecommand \bibnamefont  [1]{#1}%
\providecommand \bibfnamefont [1]{#1}%
\providecommand \citenamefont [1]{#1}%
\providecommand \href@noop [0]{\@secondoftwo}%
\providecommand \href [0]{\begingroup \@sanitize@url \@href}%
\providecommand \@href[1]{\@@startlink{#1}\@@href}%
\providecommand \@@href[1]{\endgroup#1\@@endlink}%
\providecommand \@sanitize@url [0]{\catcode `\\12\catcode `\$12\catcode
  `\&12\catcode `\#12\catcode `\^12\catcode `\_12\catcode `\%12\relax}%
\providecommand \@@startlink[1]{}%
\providecommand \@@endlink[0]{}%
\providecommand \url  [0]{\begingroup\@sanitize@url \@url }%
\providecommand \@url [1]{\endgroup\@href {#1}{\urlprefix }}%
\providecommand \urlprefix  [0]{URL }%
\providecommand \Eprint [0]{\href }%
\providecommand \doibase [0]{https://doi.org/}%
\providecommand \selectlanguage [0]{\@gobble}%
\providecommand \bibinfo  [0]{\@secondoftwo}%
\providecommand \bibfield  [0]{\@secondoftwo}%
\providecommand \translation [1]{[#1]}%
\providecommand \BibitemOpen [0]{}%
\providecommand \bibitemStop [0]{}%
\providecommand \bibitemNoStop [0]{.\EOS\space}%
\providecommand \EOS [0]{\spacefactor3000\relax}%
\providecommand \BibitemShut  [1]{\csname bibitem#1\endcsname}%
\let\auto@bib@innerbib\@empty
\bibitem [{\citenamefont {Shelby}\ \emph {et~al.}(2003)\citenamefont {Shelby},
  \citenamefont {Lim}, \citenamefont {Kuo},\ and\ \citenamefont
  {Chiu}}]{shelby2003high}%
  \BibitemOpen
  \bibfield  {author} {\bibinfo {author} {\bibfnamefont {J.~P.}\ \bibnamefont
  {Shelby}}, \bibinfo {author} {\bibfnamefont {D.~S.}\ \bibnamefont {Lim}},
  \bibinfo {author} {\bibfnamefont {J.~S.}\ \bibnamefont {Kuo}},\ and\ \bibinfo
  {author} {\bibfnamefont {D.~T.}\ \bibnamefont {Chiu}},\ }\bibfield  {title}
  {\bibinfo {title} {High radial acceleration in microvortices},\ }\href@noop
  {} {\bibfield  {journal} {\bibinfo  {journal} {Nature}\ }\textbf {\bibinfo
  {volume} {425}},\ \bibinfo {pages} {38} (\bibinfo {year} {2003})}\BibitemShut
  {NoStop}%
\bibitem [{\citenamefont {Chiu}(2007)}]{chiu2007cellular}%
  \BibitemOpen
  \bibfield  {author} {\bibinfo {author} {\bibfnamefont {D.~T.}\ \bibnamefont
  {Chiu}},\ }\bibfield  {title} {\bibinfo {title} {Cellular manipulations in
  microvortices},\ }\href@noop {} {\bibfield  {journal} {\bibinfo  {journal}
  {Analytical and bioanalytical chemistry}\ }\textbf {\bibinfo {volume}
  {387}},\ \bibinfo {pages} {17} (\bibinfo {year} {2007})}\BibitemShut
  {NoStop}%
\bibitem [{\citenamefont {Mach}\ \emph {et~al.}(2011)\citenamefont {Mach},
  \citenamefont {Kim}, \citenamefont {Arshi}, \citenamefont {Hur},\ and\
  \citenamefont {Di~Carlo}}]{mach2011automated}%
  \BibitemOpen
  \bibfield  {author} {\bibinfo {author} {\bibfnamefont {A.~J.}\ \bibnamefont
  {Mach}}, \bibinfo {author} {\bibfnamefont {J.~H.}\ \bibnamefont {Kim}},
  \bibinfo {author} {\bibfnamefont {A.}~\bibnamefont {Arshi}}, \bibinfo
  {author} {\bibfnamefont {S.~C.}\ \bibnamefont {Hur}},\ and\ \bibinfo {author}
  {\bibfnamefont {D.}~\bibnamefont {Di~Carlo}},\ }\bibfield  {title} {\bibinfo
  {title} {Automated cellular sample preparation using a
  centrifuge-on-a-chip},\ }\href@noop {} {\bibfield  {journal} {\bibinfo
  {journal} {Lab on a Chip}\ }\textbf {\bibinfo {volume} {11}},\ \bibinfo
  {pages} {2827} (\bibinfo {year} {2011})}\BibitemShut {NoStop}%
\bibitem [{\citenamefont {Martel}\ and\ \citenamefont
  {Toner}(2012)}]{martel2012inertial}%
  \BibitemOpen
  \bibfield  {author} {\bibinfo {author} {\bibfnamefont {J.~M.}\ \bibnamefont
  {Martel}}\ and\ \bibinfo {author} {\bibfnamefont {M.}~\bibnamefont {Toner}},\
  }\bibfield  {title} {\bibinfo {title} {Inertial focusing dynamics in spiral
  microchannels},\ }\href@noop {} {\bibfield  {journal} {\bibinfo  {journal}
  {physics of fluids}\ }\textbf {\bibinfo {volume} {24}},\ \bibinfo {pages}
  {032001} (\bibinfo {year} {2012})}\BibitemShut {NoStop}%
\bibitem [{\citenamefont {Mampallil}\ \emph {et~al.}(2013)\citenamefont
  {Mampallil}, \citenamefont {Tiwari}, \citenamefont {van~den Ende},\ and\
  \citenamefont {Mugele}}]{mampallil2013sample}%
  \BibitemOpen
  \bibfield  {author} {\bibinfo {author} {\bibfnamefont {D.}~\bibnamefont
  {Mampallil}}, \bibinfo {author} {\bibfnamefont {D.}~\bibnamefont {Tiwari}},
  \bibinfo {author} {\bibfnamefont {D.}~\bibnamefont {van~den Ende}},\ and\
  \bibinfo {author} {\bibfnamefont {F.}~\bibnamefont {Mugele}},\ }\bibfield
  {title} {\bibinfo {title} {Sample preconcentration inside sessile droplets
  using electrowetting},\ }\href@noop {} {\bibfield  {journal} {\bibinfo
  {journal} {Biomicrofluidics}\ }\textbf {\bibinfo {volume} {7}},\ \bibinfo
  {pages} {044102} (\bibinfo {year} {2013})}\BibitemShut {NoStop}%
\bibitem [{\citenamefont {Darhuber}\ and\ \citenamefont
  {Troian}(2005)}]{darhuber2005principles}%
  \BibitemOpen
  \bibfield  {author} {\bibinfo {author} {\bibfnamefont {A.~A.}\ \bibnamefont
  {Darhuber}}\ and\ \bibinfo {author} {\bibfnamefont {S.~M.}\ \bibnamefont
  {Troian}},\ }\bibfield  {title} {\bibinfo {title} {Principles of microfluidic
  actuation by modulation of surface stresses},\ }\href@noop {} {\bibfield
  {journal} {\bibinfo  {journal} {Annu. Rev. Fluid Mech.}\ }\textbf {\bibinfo
  {volume} {37}},\ \bibinfo {pages} {425} (\bibinfo {year} {2005})}\BibitemShut
  {NoStop}%
\bibitem [{\citenamefont {Hu}\ and\ \citenamefont
  {Larson}(2002)}]{hu2002evaporation}%
  \BibitemOpen
  \bibfield  {author} {\bibinfo {author} {\bibfnamefont {H.}~\bibnamefont
  {Hu}}\ and\ \bibinfo {author} {\bibfnamefont {R.~G.}\ \bibnamefont
  {Larson}},\ }\bibfield  {title} {\bibinfo {title} {Evaporation of a sessile
  droplet on a substrate},\ }\href@noop {} {\bibfield  {journal} {\bibinfo
  {journal} {The Journal of Physical Chemistry B}\ }\textbf {\bibinfo {volume}
  {106}},\ \bibinfo {pages} {1334} (\bibinfo {year} {2002})}\BibitemShut
  {NoStop}%
\bibitem [{\citenamefont {Brutin}\ and\ \citenamefont
  {Starov}(2018)}]{brutin2018recent}%
  \BibitemOpen
  \bibfield  {author} {\bibinfo {author} {\bibfnamefont {D.}~\bibnamefont
  {Brutin}}\ and\ \bibinfo {author} {\bibfnamefont {V.}~\bibnamefont
  {Starov}},\ }\bibfield  {title} {\bibinfo {title} {Recent advances in droplet
  wetting and evaporation},\ }\href@noop {} {\bibfield  {journal} {\bibinfo
  {journal} {Chemical Society Reviews}\ }\textbf {\bibinfo {volume} {47}},\
  \bibinfo {pages} {558} (\bibinfo {year} {2018})}\BibitemShut {NoStop}%
\bibitem [{\citenamefont {Lenshof}\ and\ \citenamefont
  {Laurell}(2011)}]{lenshof2011emerging}%
  \BibitemOpen
  \bibfield  {author} {\bibinfo {author} {\bibfnamefont {A.}~\bibnamefont
  {Lenshof}}\ and\ \bibinfo {author} {\bibfnamefont {T.}~\bibnamefont
  {Laurell}},\ }\bibfield  {title} {\bibinfo {title} {Emerging clinical
  applications of microchip-based acoustophoresis},\ }\href@noop {} {\bibfield
  {journal} {\bibinfo  {journal} {JALA: Journal of the Association for
  Laboratory Automation}\ }\textbf {\bibinfo {volume} {16}},\ \bibinfo {pages}
  {443} (\bibinfo {year} {2011})}\BibitemShut {NoStop}%
\bibitem [{\citenamefont {Riaud}\ \emph {et~al.}(2020)\citenamefont {Riaud},
  \citenamefont {Wang}, \citenamefont {Thai},\ and\ \citenamefont
  {Taly}}]{riaud2020mechanical}%
  \BibitemOpen
  \bibfield  {author} {\bibinfo {author} {\bibfnamefont {A.}~\bibnamefont
  {Riaud}}, \bibinfo {author} {\bibfnamefont {W.}~\bibnamefont {Wang}},
  \bibinfo {author} {\bibfnamefont {A.~L.}\ \bibnamefont {Thai}},\ and\
  \bibinfo {author} {\bibfnamefont {V.}~\bibnamefont {Taly}},\ }\bibfield
  {title} {\bibinfo {title} {Mechanical characterization of cells and
  microspheres sorted by acoustophoresis with in-line resistive pulse
  sensing},\ }\href@noop {} {\bibfield  {journal} {\bibinfo  {journal}
  {Physical Review Applied}\ }\textbf {\bibinfo {volume} {13}},\ \bibinfo
  {pages} {034058} (\bibinfo {year} {2020})}\BibitemShut {NoStop}%
\bibitem [{\citenamefont {Kolesnik}\ \emph {et~al.}(2021)\citenamefont
  {Kolesnik}, \citenamefont {Xu}, \citenamefont {Lee}, \citenamefont
  {Rajagopal},\ and\ \citenamefont {Collins}}]{kolesnik2021unconventional}%
  \BibitemOpen
  \bibfield  {author} {\bibinfo {author} {\bibfnamefont {K.}~\bibnamefont
  {Kolesnik}}, \bibinfo {author} {\bibfnamefont {M.}~\bibnamefont {Xu}},
  \bibinfo {author} {\bibfnamefont {P.~V.}\ \bibnamefont {Lee}}, \bibinfo
  {author} {\bibfnamefont {V.}~\bibnamefont {Rajagopal}},\ and\ \bibinfo
  {author} {\bibfnamefont {D.~J.}\ \bibnamefont {Collins}},\ }\bibfield
  {title} {\bibinfo {title} {Unconventional acoustic approaches for localized
  and designed micromanipulation},\ }\href@noop {} {\bibfield  {journal}
  {\bibinfo  {journal} {Lab on a Chip}\ } (\bibinfo {year} {2021})}\BibitemShut
  {NoStop}%
\bibitem [{\citenamefont {Shiokawa}\ \emph {et~al.}(1989)\citenamefont
  {Shiokawa}, \citenamefont {Matsui},\ and\ \citenamefont
  {Ueda}}]{shiokawa1989liquid}%
  \BibitemOpen
  \bibfield  {author} {\bibinfo {author} {\bibfnamefont {S.}~\bibnamefont
  {Shiokawa}}, \bibinfo {author} {\bibfnamefont {Y.}~\bibnamefont {Matsui}},\
  and\ \bibinfo {author} {\bibfnamefont {T.}~\bibnamefont {Ueda}},\ }\bibfield
  {title} {\bibinfo {title} {Liquid streaming and droplet formation caused by
  leaky rayleigh waves},\ }in\ \href@noop {} {\emph {\bibinfo {booktitle}
  {Proceedings., IEEE Ultrasonics Symposium,}}}\ (\bibinfo {organization}
  {IEEE},\ \bibinfo {year} {1989})\ pp.\ \bibinfo {pages}
  {643--646}\BibitemShut {NoStop}%
\bibitem [{\citenamefont {Thomas}\ and\ \citenamefont
  {Marchiano}(2003)}]{thomas2003pseudo}%
  \BibitemOpen
  \bibfield  {author} {\bibinfo {author} {\bibfnamefont {J.-L.}\ \bibnamefont
  {Thomas}}\ and\ \bibinfo {author} {\bibfnamefont {R.}~\bibnamefont
  {Marchiano}},\ }\bibfield  {title} {\bibinfo {title} {Pseudo angular momentum
  and topological charge conservation for nonlinear acoustical vortices},\
  }\href@noop {} {\bibfield  {journal} {\bibinfo  {journal} {Physical review
  letters}\ }\textbf {\bibinfo {volume} {91}},\ \bibinfo {pages} {244302}
  (\bibinfo {year} {2003})}\BibitemShut {NoStop}%
\bibitem [{\citenamefont {Lighthill}(1978)}]{lighthill1978acoustic}%
  \BibitemOpen
  \bibfield  {author} {\bibinfo {author} {\bibfnamefont {J.}~\bibnamefont
  {Lighthill}},\ }\bibfield  {title} {\bibinfo {title} {Acoustic streaming},\
  }\href@noop {} {\bibfield  {journal} {\bibinfo  {journal} {Journal of sound
  and vibration}\ }\textbf {\bibinfo {volume} {61}},\ \bibinfo {pages} {391}
  (\bibinfo {year} {1978})}\BibitemShut {NoStop}%
\bibitem [{\citenamefont {Eckart}(1948)}]{eckart1948vortices}%
  \BibitemOpen
  \bibfield  {author} {\bibinfo {author} {\bibfnamefont {C.}~\bibnamefont
  {Eckart}},\ }\bibfield  {title} {\bibinfo {title} {Vortices and streams
  caused by sound waves},\ }\href@noop {} {\bibfield  {journal} {\bibinfo
  {journal} {Physical review}\ }\textbf {\bibinfo {volume} {73}},\ \bibinfo
  {pages} {68} (\bibinfo {year} {1948})}\BibitemShut {NoStop}%
\bibitem [{\citenamefont {Riaud}\ \emph
  {et~al.}(2017{\natexlab{a}})\citenamefont {Riaud}, \citenamefont {Baudoin},
  \citenamefont {Matar}, \citenamefont {Thomas},\ and\ \citenamefont
  {Brunet}}]{riaud2017influence}%
  \BibitemOpen
  \bibfield  {author} {\bibinfo {author} {\bibfnamefont {A.}~\bibnamefont
  {Riaud}}, \bibinfo {author} {\bibfnamefont {M.}~\bibnamefont {Baudoin}},
  \bibinfo {author} {\bibfnamefont {O.~B.}\ \bibnamefont {Matar}}, \bibinfo
  {author} {\bibfnamefont {J.-L.}\ \bibnamefont {Thomas}},\ and\ \bibinfo
  {author} {\bibfnamefont {P.}~\bibnamefont {Brunet}},\ }\bibfield  {title}
  {\bibinfo {title} {On the influence of viscosity and caustics on acoustic
  streaming in sessile droplets: an experimental and a numerical study with a
  cost-effective method},\ }\href@noop {} {\bibfield  {journal} {\bibinfo
  {journal} {Journal of Fluid Mechanics}\ }\textbf {\bibinfo {volume} {821}},\
  \bibinfo {pages} {384} (\bibinfo {year} {2017}{\natexlab{a}})}\BibitemShut
  {NoStop}%
\bibitem [{\citenamefont {Li}\ \emph {et~al.}(2007)\citenamefont {Li},
  \citenamefont {Friend},\ and\ \citenamefont {Yeo}}]{li2007surface}%
  \BibitemOpen
  \bibfield  {author} {\bibinfo {author} {\bibfnamefont {H.}~\bibnamefont
  {Li}}, \bibinfo {author} {\bibfnamefont {J.~R.}\ \bibnamefont {Friend}},\
  and\ \bibinfo {author} {\bibfnamefont {L.~Y.}\ \bibnamefont {Yeo}},\
  }\bibfield  {title} {\bibinfo {title} {Surface acoustic wave concentration of
  particle and bioparticle suspensions},\ }\href@noop {} {\bibfield  {journal}
  {\bibinfo  {journal} {Biomedical microdevices}\ }\textbf {\bibinfo {volume}
  {9}},\ \bibinfo {pages} {647} (\bibinfo {year} {2007})}\BibitemShut {NoStop}%
\bibitem [{\citenamefont {Raghavan}\ \emph {et~al.}(2010)\citenamefont
  {Raghavan}, \citenamefont {Friend},\ and\ \citenamefont
  {Yeo}}]{raghavan2010particle}%
  \BibitemOpen
  \bibfield  {author} {\bibinfo {author} {\bibfnamefont {R.~V.}\ \bibnamefont
  {Raghavan}}, \bibinfo {author} {\bibfnamefont {J.~R.}\ \bibnamefont
  {Friend}},\ and\ \bibinfo {author} {\bibfnamefont {L.~Y.}\ \bibnamefont
  {Yeo}},\ }\bibfield  {title} {\bibinfo {title} {Particle concentration via
  acoustically driven microcentrifugation: micropiv flow visualization and
  numerical modelling studies},\ }\href@noop {} {\bibfield  {journal} {\bibinfo
   {journal} {Microfluidics and Nanofluidics}\ }\textbf {\bibinfo {volume}
  {8}},\ \bibinfo {pages} {73} (\bibinfo {year} {2010})}\BibitemShut {NoStop}%
\bibitem [{\citenamefont {Einstein}(1926)}]{einstein1926cause}%
  \BibitemOpen
  \bibfield  {author} {\bibinfo {author} {\bibfnamefont {A.}~\bibnamefont
  {Einstein}},\ }\bibfield  {title} {\bibinfo {title} {The cause of the
  formation of meanders in the courses of rivers and of the so-called baer’s
  law},\ }\href@noop {} {\bibfield  {journal} {\bibinfo  {journal} {Die
  Naturwissenschaften}\ }\textbf {\bibinfo {volume} {14}},\ \bibinfo {pages}
  {223} (\bibinfo {year} {1926})}\BibitemShut {NoStop}%
\bibitem [{\citenamefont {Bourquin}\ \emph {et~al.}(2014)\citenamefont
  {Bourquin}, \citenamefont {Syed}, \citenamefont {Reboud}, \citenamefont
  {Ranford-Cartwright}, \citenamefont {Barrett},\ and\ \citenamefont
  {Cooper}}]{bourquin2014rare}%
  \BibitemOpen
  \bibfield  {author} {\bibinfo {author} {\bibfnamefont {Y.}~\bibnamefont
  {Bourquin}}, \bibinfo {author} {\bibfnamefont {A.}~\bibnamefont {Syed}},
  \bibinfo {author} {\bibfnamefont {J.}~\bibnamefont {Reboud}}, \bibinfo
  {author} {\bibfnamefont {L.~C.}\ \bibnamefont {Ranford-Cartwright}}, \bibinfo
  {author} {\bibfnamefont {M.~P.}\ \bibnamefont {Barrett}},\ and\ \bibinfo
  {author} {\bibfnamefont {J.~M.}\ \bibnamefont {Cooper}},\ }\bibfield  {title}
  {\bibinfo {title} {Rare-cell enrichment by a rapid, label-free, ultrasonic
  isopycnic technique for medical diagnostics},\ }\href@noop {} {\bibfield
  {journal} {\bibinfo  {journal} {Angewandte Chemie International Edition}\
  }\textbf {\bibinfo {volume} {53}},\ \bibinfo {pages} {5587} (\bibinfo {year}
  {2014})}\BibitemShut {NoStop}%
\bibitem [{\citenamefont {Shilton}\ \emph {et~al.}(2008)\citenamefont
  {Shilton}, \citenamefont {Tan}, \citenamefont {Yeo},\ and\ \citenamefont
  {Friend}}]{shilton2008particle}%
  \BibitemOpen
  \bibfield  {author} {\bibinfo {author} {\bibfnamefont {R.}~\bibnamefont
  {Shilton}}, \bibinfo {author} {\bibfnamefont {M.~K.}\ \bibnamefont {Tan}},
  \bibinfo {author} {\bibfnamefont {L.~Y.}\ \bibnamefont {Yeo}},\ and\ \bibinfo
  {author} {\bibfnamefont {J.~R.}\ \bibnamefont {Friend}},\ }\bibfield  {title}
  {\bibinfo {title} {Particle concentration and mixing in microdrops driven by
  focused surface acoustic waves},\ }\href@noop {} {\bibfield  {journal}
  {\bibinfo  {journal} {Journal of Applied Physics}\ }\textbf {\bibinfo
  {volume} {104}},\ \bibinfo {pages} {014910} (\bibinfo {year}
  {2008})}\BibitemShut {NoStop}%
\bibitem [{\citenamefont {Zhang}\ \emph {et~al.}(2021)\citenamefont {Zhang},
  \citenamefont {Zuniga-Hertz}, \citenamefont {Zhang}, \citenamefont {Gopesh},
  \citenamefont {Fannon}, \citenamefont {Wang}, \citenamefont {Wen},
  \citenamefont {Patel},\ and\ \citenamefont {Friend}}]{zhang2021microliter}%
  \BibitemOpen
  \bibfield  {author} {\bibinfo {author} {\bibfnamefont {N.}~\bibnamefont
  {Zhang}}, \bibinfo {author} {\bibfnamefont {J.~P.}\ \bibnamefont
  {Zuniga-Hertz}}, \bibinfo {author} {\bibfnamefont {E.~Y.}\ \bibnamefont
  {Zhang}}, \bibinfo {author} {\bibfnamefont {T.}~\bibnamefont {Gopesh}},
  \bibinfo {author} {\bibfnamefont {M.~J.}\ \bibnamefont {Fannon}}, \bibinfo
  {author} {\bibfnamefont {J.}~\bibnamefont {Wang}}, \bibinfo {author}
  {\bibfnamefont {Y.}~\bibnamefont {Wen}}, \bibinfo {author} {\bibfnamefont
  {H.~H.}\ \bibnamefont {Patel}},\ and\ \bibinfo {author} {\bibfnamefont
  {J.}~\bibnamefont {Friend}},\ }\bibfield  {title} {\bibinfo {title}
  {Microliter ultrafast centrifuge platform for size-based particle and cell
  separation and extraction using novel omnidirectional spiral surface acoustic
  waves},\ }\href@noop {} {\bibfield  {journal} {\bibinfo  {journal} {Lab on a
  Chip}\ }\textbf {\bibinfo {volume} {21}},\ \bibinfo {pages} {904} (\bibinfo
  {year} {2021})}\BibitemShut {NoStop}%
\bibitem [{\citenamefont {Sudeepthi}\ \emph {et~al.}(2019)\citenamefont
  {Sudeepthi}, \citenamefont {Sen},\ and\ \citenamefont
  {Yeo}}]{sudeepthi2019aggregation}%
  \BibitemOpen
  \bibfield  {author} {\bibinfo {author} {\bibfnamefont {A.}~\bibnamefont
  {Sudeepthi}}, \bibinfo {author} {\bibfnamefont {A.~K.}\ \bibnamefont {Sen}},\
  and\ \bibinfo {author} {\bibfnamefont {L.}~\bibnamefont {Yeo}},\ }\bibfield
  {title} {\bibinfo {title} {Aggregation of a dense suspension of particles in
  a microwell using surface acoustic wave microcentrifugation},\ }\href@noop {}
  {\bibfield  {journal} {\bibinfo  {journal} {Microfluidics and Nanofluidics}\
  }\textbf {\bibinfo {volume} {23}},\ \bibinfo {pages} {1} (\bibinfo {year}
  {2019})}\BibitemShut {NoStop}%
\bibitem [{\citenamefont {Rogers}\ \emph {et~al.}(2010)\citenamefont {Rogers},
  \citenamefont {Friend},\ and\ \citenamefont {Yeo}}]{rogers2010exploitation}%
  \BibitemOpen
  \bibfield  {author} {\bibinfo {author} {\bibfnamefont {P.~R.}\ \bibnamefont
  {Rogers}}, \bibinfo {author} {\bibfnamefont {J.~R.}\ \bibnamefont {Friend}},\
  and\ \bibinfo {author} {\bibfnamefont {L.~Y.}\ \bibnamefont {Yeo}},\
  }\bibfield  {title} {\bibinfo {title} {Exploitation of surface acoustic waves
  to drive size-dependent microparticle concentration within a droplet},\
  }\href@noop {} {\bibfield  {journal} {\bibinfo  {journal} {Lab on a Chip}\
  }\textbf {\bibinfo {volume} {10}},\ \bibinfo {pages} {2979} (\bibinfo {year}
  {2010})}\BibitemShut {NoStop}%
\bibitem [{\citenamefont {Destgeer}\ \emph {et~al.}(2016)\citenamefont
  {Destgeer}, \citenamefont {Cho}, \citenamefont {Ha}, \citenamefont {Jung},
  \citenamefont {Park},\ and\ \citenamefont
  {Sung}}]{destgeer2016acoustofluidic}%
  \BibitemOpen
  \bibfield  {author} {\bibinfo {author} {\bibfnamefont {G.}~\bibnamefont
  {Destgeer}}, \bibinfo {author} {\bibfnamefont {H.}~\bibnamefont {Cho}},
  \bibinfo {author} {\bibfnamefont {B.~H.}\ \bibnamefont {Ha}}, \bibinfo
  {author} {\bibfnamefont {J.~H.}\ \bibnamefont {Jung}}, \bibinfo {author}
  {\bibfnamefont {J.}~\bibnamefont {Park}},\ and\ \bibinfo {author}
  {\bibfnamefont {H.~J.}\ \bibnamefont {Sung}},\ }\bibfield  {title} {\bibinfo
  {title} {Acoustofluidic particle manipulation inside a sessile droplet: four
  distinct regimes of particle concentration},\ }\href@noop {} {\bibfield
  {journal} {\bibinfo  {journal} {Lab on a Chip}\ }\textbf {\bibinfo {volume}
  {16}},\ \bibinfo {pages} {660} (\bibinfo {year} {2016})}\BibitemShut
  {NoStop}%
\bibitem [{\citenamefont {Destgeer}\ \emph {et~al.}(2017)\citenamefont
  {Destgeer}, \citenamefont {Jung}, \citenamefont {Park}, \citenamefont
  {Ahmed},\ and\ \citenamefont {Sung}}]{destgeer2017particle}%
  \BibitemOpen
  \bibfield  {author} {\bibinfo {author} {\bibfnamefont {G.}~\bibnamefont
  {Destgeer}}, \bibinfo {author} {\bibfnamefont {J.~H.}\ \bibnamefont {Jung}},
  \bibinfo {author} {\bibfnamefont {J.}~\bibnamefont {Park}}, \bibinfo {author}
  {\bibfnamefont {H.}~\bibnamefont {Ahmed}},\ and\ \bibinfo {author}
  {\bibfnamefont {H.~J.}\ \bibnamefont {Sung}},\ }\bibfield  {title} {\bibinfo
  {title} {Particle separation inside a sessile droplet with variable contact
  angle using surface acoustic waves},\ }\href@noop {} {\bibfield  {journal}
  {\bibinfo  {journal} {Analytical chemistry}\ }\textbf {\bibinfo {volume}
  {89}},\ \bibinfo {pages} {736} (\bibinfo {year} {2017})}\BibitemShut
  {NoStop}%
\bibitem [{\citenamefont {Gu}\ \emph {et~al.}(2021)\citenamefont {Gu},
  \citenamefont {Chen}, \citenamefont {Mao}, \citenamefont {Bachman},
  \citenamefont {Becker}, \citenamefont {Rufo}, \citenamefont {Wang},
  \citenamefont {Zhang}, \citenamefont {Mai}, \citenamefont {Yang} \emph
  {et~al.}}]{gu2021acoustofluidic}%
  \BibitemOpen
  \bibfield  {author} {\bibinfo {author} {\bibfnamefont {Y.}~\bibnamefont
  {Gu}}, \bibinfo {author} {\bibfnamefont {C.}~\bibnamefont {Chen}}, \bibinfo
  {author} {\bibfnamefont {Z.}~\bibnamefont {Mao}}, \bibinfo {author}
  {\bibfnamefont {H.}~\bibnamefont {Bachman}}, \bibinfo {author} {\bibfnamefont
  {R.}~\bibnamefont {Becker}}, \bibinfo {author} {\bibfnamefont
  {J.}~\bibnamefont {Rufo}}, \bibinfo {author} {\bibfnamefont {Z.}~\bibnamefont
  {Wang}}, \bibinfo {author} {\bibfnamefont {P.}~\bibnamefont {Zhang}},
  \bibinfo {author} {\bibfnamefont {J.}~\bibnamefont {Mai}}, \bibinfo {author}
  {\bibfnamefont {S.}~\bibnamefont {Yang}}, \emph {et~al.},\ }\bibfield
  {title} {\bibinfo {title} {Acoustofluidic centrifuge for nanoparticle
  enrichment and separation},\ }\href@noop {} {\bibfield  {journal} {\bibinfo
  {journal} {Science advances}\ }\textbf {\bibinfo {volume} {7}},\ \bibinfo
  {pages} {eabc0467} (\bibinfo {year} {2021})}\BibitemShut {NoStop}%
\bibitem [{\citenamefont {Zhang}\ \emph {et~al.}(2009)\citenamefont {Zhang},
  \citenamefont {Liu}, \citenamefont {Jiang},\ and\ \citenamefont
  {Fei}}]{zhang2009rapid}%
  \BibitemOpen
  \bibfield  {author} {\bibinfo {author} {\bibfnamefont {A.}~\bibnamefont
  {Zhang}}, \bibinfo {author} {\bibfnamefont {W.}~\bibnamefont {Liu}}, \bibinfo
  {author} {\bibfnamefont {Z.}~\bibnamefont {Jiang}},\ and\ \bibinfo {author}
  {\bibfnamefont {J.}~\bibnamefont {Fei}},\ }\bibfield  {title} {\bibinfo
  {title} {Rapid concentration of particle and bioparticle suspension based on
  surface acoustic wave},\ }\href@noop {} {\bibfield  {journal} {\bibinfo
  {journal} {Applied Acoustics}\ }\textbf {\bibinfo {volume} {70}},\ \bibinfo
  {pages} {1137} (\bibinfo {year} {2009})}\BibitemShut {NoStop}%
\bibitem [{\citenamefont {Riaud}\ \emph
  {et~al.}(2015{\natexlab{a}})\citenamefont {Riaud}, \citenamefont {Thomas},
  \citenamefont {Charron}, \citenamefont {Bussonni{\`e}re}, \citenamefont
  {Matar},\ and\ \citenamefont {Baudoin}}]{riaud2015anisotropic}%
  \BibitemOpen
  \bibfield  {author} {\bibinfo {author} {\bibfnamefont {A.}~\bibnamefont
  {Riaud}}, \bibinfo {author} {\bibfnamefont {J.-L.}\ \bibnamefont {Thomas}},
  \bibinfo {author} {\bibfnamefont {E.}~\bibnamefont {Charron}}, \bibinfo
  {author} {\bibfnamefont {A.}~\bibnamefont {Bussonni{\`e}re}}, \bibinfo
  {author} {\bibfnamefont {O.~B.}\ \bibnamefont {Matar}},\ and\ \bibinfo
  {author} {\bibfnamefont {M.}~\bibnamefont {Baudoin}},\ }\bibfield  {title}
  {\bibinfo {title} {Anisotropic swirling surface acoustic waves from inverse
  filtering for on-chip generation of acoustic vortices},\ }\href@noop {}
  {\bibfield  {journal} {\bibinfo  {journal} {Physical Review Applied}\
  }\textbf {\bibinfo {volume} {4}},\ \bibinfo {pages} {034004} (\bibinfo {year}
  {2015}{\natexlab{a}})}\BibitemShut {NoStop}%
\bibitem [{\citenamefont {Riaud}\ \emph
  {et~al.}(2015{\natexlab{b}})\citenamefont {Riaud}, \citenamefont {Thomas},
  \citenamefont {Baudoin},\ and\ \citenamefont {Matar}}]{riaud2015taming}%
  \BibitemOpen
  \bibfield  {author} {\bibinfo {author} {\bibfnamefont {A.}~\bibnamefont
  {Riaud}}, \bibinfo {author} {\bibfnamefont {J.-L.}\ \bibnamefont {Thomas}},
  \bibinfo {author} {\bibfnamefont {M.}~\bibnamefont {Baudoin}},\ and\ \bibinfo
  {author} {\bibfnamefont {O.~B.}\ \bibnamefont {Matar}},\ }\bibfield  {title}
  {\bibinfo {title} {Taming the degeneration of bessel beams at an
  anisotropic-isotropic interface: Toward three-dimensional control of confined
  vortical waves},\ }\href@noop {} {\bibfield  {journal} {\bibinfo  {journal}
  {Physical Review E}\ }\textbf {\bibinfo {volume} {92}},\ \bibinfo {pages}
  {063201} (\bibinfo {year} {2015}{\natexlab{b}})}\BibitemShut {NoStop}%
\bibitem [{\citenamefont {Riaud}\ \emph
  {et~al.}(2017{\natexlab{b}})\citenamefont {Riaud}, \citenamefont {Baudoin},
  \citenamefont {Matar}, \citenamefont {Becerra},\ and\ \citenamefont
  {Thomas}}]{riaud2017selective}%
  \BibitemOpen
  \bibfield  {author} {\bibinfo {author} {\bibfnamefont {A.}~\bibnamefont
  {Riaud}}, \bibinfo {author} {\bibfnamefont {M.}~\bibnamefont {Baudoin}},
  \bibinfo {author} {\bibfnamefont {O.~B.}\ \bibnamefont {Matar}}, \bibinfo
  {author} {\bibfnamefont {L.}~\bibnamefont {Becerra}},\ and\ \bibinfo {author}
  {\bibfnamefont {J.-L.}\ \bibnamefont {Thomas}},\ }\bibfield  {title}
  {\bibinfo {title} {Selective manipulation of microscopic particles with
  precursor swirling rayleigh waves},\ }\href@noop {} {\bibfield  {journal}
  {\bibinfo  {journal} {Physical Review Applied}\ }\textbf {\bibinfo {volume}
  {7}},\ \bibinfo {pages} {024007} (\bibinfo {year}
  {2017}{\natexlab{b}})}\BibitemShut {NoStop}%
\bibitem [{\citenamefont {Baresch}\ \emph {et~al.}(2013)\citenamefont
  {Baresch}, \citenamefont {Thomas},\ and\ \citenamefont
  {Marchiano}}]{baresch2013three}%
  \BibitemOpen
  \bibfield  {author} {\bibinfo {author} {\bibfnamefont {D.}~\bibnamefont
  {Baresch}}, \bibinfo {author} {\bibfnamefont {J.-L.}\ \bibnamefont
  {Thomas}},\ and\ \bibinfo {author} {\bibfnamefont {R.}~\bibnamefont
  {Marchiano}},\ }\bibfield  {title} {\bibinfo {title} {Three-dimensional
  acoustic radiation force on an arbitrarily located elastic sphere},\
  }\href@noop {} {\bibfield  {journal} {\bibinfo  {journal} {The Journal of the
  Acoustical Society of America}\ }\textbf {\bibinfo {volume} {133}},\ \bibinfo
  {pages} {25} (\bibinfo {year} {2013})}\BibitemShut {NoStop}%
\bibitem [{\citenamefont {Gong}\ and\ \citenamefont
  {Baudoin}(2021)}]{gong2021equivalence}%
  \BibitemOpen
  \bibfield  {author} {\bibinfo {author} {\bibfnamefont {Z.}~\bibnamefont
  {Gong}}\ and\ \bibinfo {author} {\bibfnamefont {M.}~\bibnamefont {Baudoin}},\
  }\bibfield  {title} {\bibinfo {title} {Equivalence between angular
  spectrum-based and multipole expansion-based formulas of the acoustic
  radiation force and torque},\ }\href@noop {} {\bibfield  {journal} {\bibinfo
  {journal} {The Journal of the Acoustical Society of America}\ }\textbf
  {\bibinfo {volume} {149}},\ \bibinfo {pages} {3469} (\bibinfo {year}
  {2021})}\BibitemShut {NoStop}%
\bibitem [{\citenamefont {Doinikov}(1994)}]{doinikov1994acoustic}%
  \BibitemOpen
  \bibfield  {author} {\bibinfo {author} {\bibfnamefont {A.}~\bibnamefont
  {Doinikov}},\ }\bibfield  {title} {\bibinfo {title} {Acoustic radiation
  pressure on a rigid sphere in a viscous fluid},\ }\href@noop {} {\bibfield
  {journal} {\bibinfo  {journal} {Proceedings of the Royal Society of London.
  Series A: Mathematical and Physical Sciences}\ }\textbf {\bibinfo {volume}
  {447}},\ \bibinfo {pages} {447} (\bibinfo {year} {1994})}\BibitemShut
  {NoStop}%
\bibitem [{\citenamefont {Baasch}\ \emph {et~al.}(2019)\citenamefont {Baasch},
  \citenamefont {Pavlic},\ and\ \citenamefont {Dual}}]{baasch2019acoustic}%
  \BibitemOpen
  \bibfield  {author} {\bibinfo {author} {\bibfnamefont {T.}~\bibnamefont
  {Baasch}}, \bibinfo {author} {\bibfnamefont {A.}~\bibnamefont {Pavlic}},\
  and\ \bibinfo {author} {\bibfnamefont {J.}~\bibnamefont {Dual}},\ }\bibfield
  {title} {\bibinfo {title} {Acoustic radiation force acting on a heavy
  particle in a standing wave can be dominated by the acoustic
  microstreaming},\ }\href@noop {} {\bibfield  {journal} {\bibinfo  {journal}
  {Physical Review E}\ }\textbf {\bibinfo {volume} {100}},\ \bibinfo {pages}
  {061102} (\bibinfo {year} {2019})}\BibitemShut {NoStop}%
\bibitem [{\citenamefont {Strobl}\ \emph {et~al.}(2004)\citenamefont {Strobl},
  \citenamefont {Sch{\"a}flein}, \citenamefont {Beierlein}, \citenamefont
  {Ebbecke},\ and\ \citenamefont {Wixforth}}]{strobl2004carbon}%
  \BibitemOpen
  \bibfield  {author} {\bibinfo {author} {\bibfnamefont {C.~J.}\ \bibnamefont
  {Strobl}}, \bibinfo {author} {\bibfnamefont {C.}~\bibnamefont
  {Sch{\"a}flein}}, \bibinfo {author} {\bibfnamefont {U.}~\bibnamefont
  {Beierlein}}, \bibinfo {author} {\bibfnamefont {J.}~\bibnamefont {Ebbecke}},\
  and\ \bibinfo {author} {\bibfnamefont {A.}~\bibnamefont {Wixforth}},\
  }\bibfield  {title} {\bibinfo {title} {Carbon nanotube alignment by surface
  acoustic waves},\ }\href@noop {} {\bibfield  {journal} {\bibinfo  {journal}
  {Applied physics letters}\ }\textbf {\bibinfo {volume} {85}},\ \bibinfo
  {pages} {1427} (\bibinfo {year} {2004})}\BibitemShut {NoStop}%
\bibitem [{\citenamefont {Ma}\ \emph {et~al.}(2015)\citenamefont {Ma},
  \citenamefont {Guo}, \citenamefont {Liu},\ and\ \citenamefont
  {Ai}}]{ma2015patterning}%
  \BibitemOpen
  \bibfield  {author} {\bibinfo {author} {\bibfnamefont {Z.}~\bibnamefont
  {Ma}}, \bibinfo {author} {\bibfnamefont {J.}~\bibnamefont {Guo}}, \bibinfo
  {author} {\bibfnamefont {Y.~J.}\ \bibnamefont {Liu}},\ and\ \bibinfo {author}
  {\bibfnamefont {Y.}~\bibnamefont {Ai}},\ }\bibfield  {title} {\bibinfo
  {title} {The patterning mechanism of carbon nanotubes using surface acoustic
  waves: The acoustic radiation effect or the dielectrophoretic effect},\
  }\href@noop {} {\bibfield  {journal} {\bibinfo  {journal} {Nanoscale}\
  }\textbf {\bibinfo {volume} {7}},\ \bibinfo {pages} {14047} (\bibinfo {year}
  {2015})}\BibitemShut {NoStop}%
\bibitem [{\citenamefont {White}\ and\ \citenamefont
  {Voltmer}(1965)}]{white1965direct}%
  \BibitemOpen
  \bibfield  {author} {\bibinfo {author} {\bibfnamefont {R.~M.}\ \bibnamefont
  {White}}\ and\ \bibinfo {author} {\bibfnamefont {F.~W.}\ \bibnamefont
  {Voltmer}},\ }\bibfield  {title} {\bibinfo {title} {Direct piezoelectric
  coupling to surface elastic waves},\ }\href@noop {} {\bibfield  {journal}
  {\bibinfo  {journal} {Applied physics letters}\ }\textbf {\bibinfo {volume}
  {7}},\ \bibinfo {pages} {314} (\bibinfo {year} {1965})}\BibitemShut {NoStop}%
\bibitem [{\citenamefont {Bleustein}(1968)}]{bleustein1968new}%
  \BibitemOpen
  \bibfield  {author} {\bibinfo {author} {\bibfnamefont {J.~L.}\ \bibnamefont
  {Bleustein}},\ }\bibfield  {title} {\bibinfo {title} {A new surface wave in
  piezoelectric materials},\ }\href@noop {} {\bibfield  {journal} {\bibinfo
  {journal} {Applied Physics Letters}\ }\textbf {\bibinfo {volume} {13}},\
  \bibinfo {pages} {412} (\bibinfo {year} {1968})}\BibitemShut {NoStop}%
\bibitem [{\citenamefont {Stone}(2000)}]{stone2000philip}%
  \BibitemOpen
  \bibfield  {author} {\bibinfo {author} {\bibfnamefont {H.~A.}\ \bibnamefont
  {Stone}},\ }\bibfield  {title} {\bibinfo {title} {Philip saffman and viscous
  flow theory},\ }\href@noop {} {\bibfield  {journal} {\bibinfo  {journal}
  {Journal of Fluid Mechanics}\ }\textbf {\bibinfo {volume} {409}},\ \bibinfo
  {pages} {165} (\bibinfo {year} {2000})}\BibitemShut {NoStop}%
\bibitem [{\citenamefont {Goldman}\ \emph {et~al.}(1967)\citenamefont
  {Goldman}, \citenamefont {Cox},\ and\ \citenamefont
  {Brenner}}]{goldman1967slow}%
  \BibitemOpen
  \bibfield  {author} {\bibinfo {author} {\bibfnamefont {A.}~\bibnamefont
  {Goldman}}, \bibinfo {author} {\bibfnamefont {R.~G.}\ \bibnamefont {Cox}},\
  and\ \bibinfo {author} {\bibfnamefont {H.}~\bibnamefont {Brenner}},\
  }\bibfield  {title} {\bibinfo {title} {Slow viscous motion of a sphere
  parallel to a plane wall—ii couette flow},\ }\href@noop {} {\bibfield
  {journal} {\bibinfo  {journal} {Chemical engineering science}\ }\textbf
  {\bibinfo {volume} {22}},\ \bibinfo {pages} {653} (\bibinfo {year}
  {1967})}\BibitemShut {NoStop}%
\bibitem [{\citenamefont {Krishnan}\ and\ \citenamefont
  {Leighton~Jr}(1995)}]{krishnan1995inertial}%
  \BibitemOpen
  \bibfield  {author} {\bibinfo {author} {\bibfnamefont {G.~P.}\ \bibnamefont
  {Krishnan}}\ and\ \bibinfo {author} {\bibfnamefont {D.~T.}\ \bibnamefont
  {Leighton~Jr}},\ }\bibfield  {title} {\bibinfo {title} {Inertial lift on a
  moving sphere in contact with a plane wall in a shear flow},\ }\href@noop {}
  {\bibfield  {journal} {\bibinfo  {journal} {Physics of Fluids}\ }\textbf
  {\bibinfo {volume} {7}},\ \bibinfo {pages} {2538} (\bibinfo {year}
  {1995})}\BibitemShut {NoStop}%
\bibitem [{\citenamefont {Shields}(1936)}]{shields1936anwendung}%
  \BibitemOpen
  \bibfield  {author} {\bibinfo {author} {\bibfnamefont {A.}~\bibnamefont
  {Shields}},\ }\bibfield  {title} {\bibinfo {title} {Anwendung der
  aehnlichkeitsmechanik und der turbulenzforschung auf die geschiebebewegung},\
  }\href@noop {} {\bibfield  {journal} {\bibinfo  {journal} {PhD Thesis
  Technical University Berlin}\ } (\bibinfo {year} {1936})}\BibitemShut
  {NoStop}%
\bibitem [{\citenamefont {Agudo}\ \emph {et~al.}(2017)\citenamefont {Agudo},
  \citenamefont {Illigmann}, \citenamefont {Luzi}, \citenamefont {Laukart},
  \citenamefont {Delgado},\ and\ \citenamefont {Wierschem}}]{agudo2017shear}%
  \BibitemOpen
  \bibfield  {author} {\bibinfo {author} {\bibfnamefont {J.}~\bibnamefont
  {Agudo}}, \bibinfo {author} {\bibfnamefont {C.}~\bibnamefont {Illigmann}},
  \bibinfo {author} {\bibfnamefont {G.}~\bibnamefont {Luzi}}, \bibinfo {author}
  {\bibfnamefont {A.}~\bibnamefont {Laukart}}, \bibinfo {author} {\bibfnamefont
  {A.}~\bibnamefont {Delgado}},\ and\ \bibinfo {author} {\bibfnamefont
  {A.}~\bibnamefont {Wierschem}},\ }\bibfield  {title} {\bibinfo {title}
  {Shear-induced incipient motion of a single sphere on uniform substrates at
  low particle reynolds numbers},\ }\href@noop {} {\bibfield  {journal}
  {\bibinfo  {journal} {Journal of Fluid Mechanics}\ }\textbf {\bibinfo
  {volume} {825}},\ \bibinfo {pages} {284} (\bibinfo {year}
  {2017})}\BibitemShut {NoStop}%
\bibitem [{\citenamefont {Topic}\ \emph {et~al.}(2019)\citenamefont {Topic},
  \citenamefont {Retzepoglu}, \citenamefont {Wensing}, \citenamefont
  {Illigmann}, \citenamefont {Luzi}, \citenamefont {Agudo},\ and\ \citenamefont
  {Wierschem}}]{topic2019effect}%
  \BibitemOpen
  \bibfield  {author} {\bibinfo {author} {\bibfnamefont {N.}~\bibnamefont
  {Topic}}, \bibinfo {author} {\bibfnamefont {S.}~\bibnamefont {Retzepoglu}},
  \bibinfo {author} {\bibfnamefont {M.}~\bibnamefont {Wensing}}, \bibinfo
  {author} {\bibfnamefont {C.}~\bibnamefont {Illigmann}}, \bibinfo {author}
  {\bibfnamefont {G.}~\bibnamefont {Luzi}}, \bibinfo {author} {\bibfnamefont
  {J.}~\bibnamefont {Agudo}},\ and\ \bibinfo {author} {\bibfnamefont
  {A.}~\bibnamefont {Wierschem}},\ }\bibfield  {title} {\bibinfo {title}
  {Effect of particle size ratio on shear-induced onset of particle motion at
  low particle reynolds numbers: From high shielding to roughness},\
  }\href@noop {} {\bibfield  {journal} {\bibinfo  {journal} {Physics of
  Fluids}\ }\textbf {\bibinfo {volume} {31}},\ \bibinfo {pages} {063305}
  (\bibinfo {year} {2019})}\BibitemShut {NoStop}%
\bibitem [{\citenamefont {Doinikov}\ \emph {et~al.}(2020)\citenamefont
  {Doinikov}, \citenamefont {Gerlt},\ and\ \citenamefont
  {Dual}}]{doinikov2020acoustic}%
  \BibitemOpen
  \bibfield  {author} {\bibinfo {author} {\bibfnamefont {A.~A.}\ \bibnamefont
  {Doinikov}}, \bibinfo {author} {\bibfnamefont {M.~S.}\ \bibnamefont
  {Gerlt}},\ and\ \bibinfo {author} {\bibfnamefont {J.}~\bibnamefont {Dual}},\
  }\bibfield  {title} {\bibinfo {title} {Acoustic radiation forces produced by
  sharp-edge structures in microfluidic systems},\ }\href@noop {} {\bibfield
  {journal} {\bibinfo  {journal} {Physical Review Letters}\ }\textbf {\bibinfo
  {volume} {124}},\ \bibinfo {pages} {154501} (\bibinfo {year}
  {2020})}\BibitemShut {NoStop}%
\end{thebibliography}%

\begin{acknowledgments}
	
This work was supported by the National Natural Science Foundation of China with Grant No. 12004078 and No. 61874033, State Key Lab of ASIC and System, Fudan University 2021MS001, 2021MS002 and 2020KF006.
	
\end{acknowledgments}

\section{Author contributions}
AR and JZ proposed the research, SS did the experiments, AR and SS did the simulations. All the authors wrote the paper together.

\section{Competing interests}
The authors declare no competing interests.

\end{document}